\documentclass[twoside,english,superscriptaddress,nofootinbib,preprintnumbers, aps]{revtex4}

\usepackage{euscript,amssymb}
\usepackage{amsthm}
\usepackage{graphicx}
\usepackage{amsfonts}
\usepackage{amsmath}
\usepackage{amssymb}
\usepackage{fancyhdr}
\usepackage{epsfig}
\usepackage[usenames,dvipsnames]{xcolor}
\usepackage{esint}
\usepackage{bm}
\usepackage[unicode=true, pdfusetitle, bookmarks=true,bookmarksnumbered=false,bookmarksopen=false, breaklinks=false,pdfborder={0 0 1},backref=false,colorlinks=false]{hyperref}

\begin{document}

\title{{QFT on quantum spacetime: a compatible classical framework}}

\author{Andrea Dapor}
\email{adapor@fuw.edu.pl} \affiliation{Instytut Fizyki Teoretycznej, 
Uniwersytet Warszawski, ul. Ho\.{z}a 69, 00-681 Warszawa, Poland}

\author{Jerzy Lewandowski}
\email{jerzy.lewandowski@fuw.edu.pl} \affiliation{Instytut Fizyki Teoretycznej, 
Uniwersytet Warszawski, ul. Ho\.{z}a 69, 00-681 Warszawa, Poland}
\affiliation{Institute for Quantum Gravity (IQG), FAU Erlangen -- Nurnberg,
Staudtstr. 7, 91058 Erlangen, Germany}

\author{Jacek Puchta}
\email{jpa@fuw.edu.pl} \affiliation{Instytut Fizyki Teoretycznej, 
Uniwersytet Warszawski, ul. Ho\.{z}a 69, 00-681 Warszawa, Poland}

\begin{abstract}
{We develop a systematic classical framework to accommodate canonical quantization of geometric and matter perturbations on a quantum homogeneous isotropic flat spacetime. The existing approach of standard cosmological perturbations is indeed proved to be good only up to first order in the inhomogeneities, and only if the background is treated classically. To consistently quantize the perturbations \emph{and} the background, a new set of classical phase space variables is required. We show that, in a natural gauge, a set of such Dirac observables exists, and their algebra is of the canonical form. Finally, we compute the physical Hamiltonian that generates the dynamics of such observables with respect to the homogeneous part of a K-G ''clock'' field $T$. The results of this work provide a good starting point to understanding and calculating effects that quantum cosmological spacetime in the background has on the quantum perturbations of the metric tensor and of matter fields.}
\end{abstract}

\date{\today}

\pacs{???}

\maketitle
\section{Introduction}
\subsection{Motivation: towards a new quantum theory}
The framework we present in this paper is classical, but the aim is introduction of quantum test fields on a quantum spacetime.  
While the theory of test quantum fields propagating on classical FRW spacetime is very well known, a theory of quantum fields on a {\it quantum} universe in the background is a scenario considered much less often. The first step in that direction was made in \cite{AKL}. 
In that work, the quantum background was provided by the loop quantum cosmology model\footnote{
For an account on loop quantum cosmology and loop quantum gravity, refer for example to \cite{zako1, zako2, LQG1, LQG2, LQC1, LQC2, LQC3}.
}
of the homogenues isotropic universe characterized by the scale factor $e^\alpha$ coupled to a homogeneus massless K-G field $T^{(0)}$ {}{(playing the role of physical time), and the test field was a second K-G field $\delta\phi$}. {}{This model was derived from the (suitably simplified) theory of two K-G fields, $T(x)$ and $\phi(x)$, coupled to gravitational field $g_{\mu \nu}(x)$}: the authors started with the scalar constraint of the full theory, expanded it around the homogeneus solutions (with $\phi=0$), and dropped all the degrees of freedom except for (i) the scale $\alpha$ of the universe, (ii) the homogeneus part $T^{(0)}$ of the first K-G field, and (iii) the perturbations $\delta\phi$ of the second K-G field. All these three remaining degrees of freedom were coupled to each other and subject to a quantum scalar constraint defined by the truncated quantum constraint operator $\hat{C}=\hat{C}_{\alpha}+\hat{C}_{T^{(0)}}+\hat{C}_{\delta \phi}=0$. (The vector constraint was satisfied automatically at the classical level in the test field approximation, therefore it was ignored at the quantum level.) That idea was later generalized to the Bianchi I quantum spacetimes \cite{DLT}.

The goal of that new approach to QFT on quantum spacetimes was to gain some insight into possible effects of the quantum nature of geometry on the propagation of test fields. A first intriguing conclusion came from the study of a mechanism of the emergence of a classical spacetime from such quantum system. The classical spacetime emerges  as a metric tensor effectively felt by the modes of the quantum test field. It is obtained from the quantum dynamics of the modes, and differs from solutions to the classical Enstein equation with suitable quantum corrections. For that reason, it was later called ``dressed'' \cite{ASH}. If the test K-G field is massless, then all its modes experience a {\it single} dressed metric, independently of their momenta. That phenomenon may be interpreted as the absence of so-called Lorenz symmetry violation. If the test K-G field is massive, on the other hand,  then the dressed metric felt by each mode depends of the direction of its momentum. In particular, the dressed metric is {\it not any longer space-isotropic}, even in the case of an isotropic quantum universe in the background \cite{iso}. One might say that, from the point of view of each mode, the isotropy is broken by quantum geometry effects! {}{These results were insensitive on possible choices of the quantum model for the homogeneous degrees of freedom, be them loop quantum cosmology \cite{LQC1, LQC2, LQC3} or Wheeler-deWitt quatum cosmology \cite{WdW1, WdW2}. However, the findings outlined above were derived using a quite crude approximation, therefore one expects more from a systematic approach.}

Since the concept of the test quantum field on the quantum cosmological background spacetime has proved to be quite fruitful, it is worth to extend it to perturbations of  the gravitational field  and to perturbations of the the scalar field $T$ present in the background spacetime. {}{An attempt to achieve that goal was made in \cite{ASH}. Therein, the approach of \cite{AKL} is reconsidered. However, it is developed in a somewhat different direction. Namely, as a starting point a very well known standard  perturbation theory of classical mathematical cosmology \cite{Lang,Stand1,Stand2} is taken. 
Each dynamical variable $\gamma$ (a coordinate in the phase space) is expanded as
\begin{align} \label{BaD}
\gamma = \tilde{\gamma}^{(0)} + \epsilon\delta \gamma^{(1)} + \frac{1}{2}\epsilon^2\delta \gamma^{(2)} + ...
\end{align}
where $\tilde{\gamma}^{(0)}$ encodes the background part. The background part is decoupled from the higher order perturbations. Thus, the 1st order perturbations $\delta \gamma^{(1)}$ gain their own phase space and are subject to a theory introduced on a fixed background $\tilde{\gamma}^{(0)}$. Such framework is powerful in cosmology and useful for example in the context of QFT on curved classical spacetime, because \emph{the background spacetime has been fixed from the beginning as a solution to the unperturbed Einstein equations}}, and the phase space of the system consists of the perturbation sector only. {}{In \cite{ASH} that framework is quantized.
Specifically, one quantizes the perturbations order by order, decoupled from the background and from each other. What we want to consider instead, is a joint quantization of the perturbations \emph{and} of the background (as is the case of \cite{AKL, DLT}). Therefore, we will propose in the current paper to go in an alternative  direction to that that of \cite{ASH}. To this end} we need to develop a classical framework that keeps the original coupling between the perturbations and the background. {}{Another  proposal for a systematic development of the theory of quantum perturbations coupled to quantum background was made in \cite{MikelGuillermoJavier}. The starting point of the authors is the classical framework of \cite{HalliwelHawking} available for the $k=1$ cosmology (a spherical universe). Their framework combines  the LQC quantization of the background with a unique quantization of scalar cosmological perturbations on a classical background with $k=1$ \cite{MikelGuillermoJavierJose}. That proposal is satisfactory from the point of view of deriving a theory of perturbations from the full theory. What we are looking for in the current paper is a similar classical starting point available for a flat universe.} A similar idea is presented in \cite{bianka}, where the authors study the problem of evolution of perturbations around the cosmological sector of general relativity in Ashtekar-Barbero variables.

%
%
%
\subsection{Comparison and contrast with the standard approach}
 {The current paper\footnote{
The first ideas were presented in \cite{jacek}.
}
provides such classical framework, suitable for the project of studying QFT on quantum spacetime. For the sake of self-sufficiency, we start from scratch. We address the full theory of gravity coupled to two K-G fields, and systematically develop a classical framework in which the perturbations and the background together set a phase space. To this end, in Section \ref{clasdeo} we construct the full phase space of the system of the two Klein-Gordon fields coupled to the gravitational field. In Section \ref{BiG} we briefly present the first step of the starndar approach of cosmological perturbation theory, namely the definition of the background variables $\tilde{\gamma}^{(0)}$ of (\ref{BaD}). We do this to fix the notation and in order to draw a comparison with our approach, which is constructed in the remainder of the paper.}

 {In Section \ref{fullcor} we introduce a global coordinate system on the full phase space, in which every field is splitted as its homogeneous and inhomogeneous parts:
\begin{align} \label{GooD}
\gamma = \gamma^{(0)} + \delta \gamma
\end{align}
Note that, contrary to (\ref{BaD}), here there are no higher order terms. This is because the expression (\ref{GooD}) is \emph{not} the first two terms of the Taylor expansion (there is no ``small parameter'' $\epsilon$, and both terms are finite), but rather an exact, unique decomposition giving rise to a certain coordinate system on the phase space.\footnote{
 {More rigorously, $\delta \gamma$ are functions on the phase space $\Gamma$ of the theory, rather then elements of the tangent space $T_{\tilde{\gamma}^{(0)}} \Gamma$ at a fixed background solution $\tilde{\gamma}^{(0)}$.}
}
To give an example, for the K-G field $T(x)$ sector, we will define $T^{(0)}$ to be the homogeneous part and $\delta T(x)$ the inhomogeneous part: that split is always well defined, and does not involve any knowledge of the dynamics (contrary to (\ref{BaD}), where  given an exact $T(x)$, the definition of $T^{(0)}$ is the part of $T(x)$ that satisfies unperturbed K-G equation). Because of the kinematical nature of this decomposition, it follows that in our sense, $\delta T(x)$ does not involve any correction to the homogeneous part of the field: it is all absorbed in $T^{(0)}$ (which in the consequence does not satisfy the unperturbed K-G equation).\footnote{
 {Actually, if one considers the dynamics up to the first order, the variables $\gamma^{(0)}$ do satisfy the Klein-Gordon-Einstein equations (with $\delta\gamma=0$), and the variables $\delta\gamma$ satisfy the linearized equations on the background $\gamma^{(0)}$. Thus, up to the $1$st order, the two approaches agree. However, it will be clear that, if quadratic order were considered, then this would not be the case, since $\gamma^{(0)}$ would know about ``back-reaction of perturbations on the background''.}
}}

 {As already said, the reader should not be mislead by the symbol $\delta$ in front of $\gamma$: at this kinematical level nothing is ``small'' yet. The next step of the programme is to consider the constraints of the full theory on the phase space thus coordinatized, and reduce them to the constraint surface. In order to do this explicitly, however, we are forced at this point to \emph{consider only those spacetimes for which $\delta \gamma$ is indeed small}. Therefore, in Section (\ref{constup}) we carry out a Taylor expansion and Fourier mode-decomposition of the full constraints as functions on the phase space expressed in terms of $\gamma = \gamma^{(0)} + \delta \gamma$, for $\delta \gamma \ll \gamma^{(0)}$. In Section \ref{solofconstruno} we study the gauge transformations generated by the constraints, which finally allow to restrict to the reduced physical phase space. In Section \ref{dinamo} we construct the Dirac observales, along with a 1-dimensional group of automorphisms of their algebra parametrized by the variable $T^{(0)}$, and find the generator of such group, i.e. the physical Hamiltonian. A comparison with the main results of standard cosmological perturbation theory is drawn in Section \ref{MSwrongs}, where we explain why Mukhanov-Sasaki variables cannot fit together with a canonical quantization of the background spacetime. Indeed, the structure needed for such canonical quantization is the Poisson algebra of the Dirac observables and the physical Hamiltonian. Not only Mukhanov-Sasaki variables are not Dirac observables at higher orders, but also they present a non-trivial commutation relation with the background variables. As we will explain in detail, this fact is due to the dynamical nature of the expansion (\ref{BaD}). On the other hand, the essential feature of our framework is that $\gamma^{(0)}$ and $\delta \gamma$ in (\ref{GooD}) are subject to the unchanged Poisson algebra of the full theory, contrary to $\tilde{\gamma}^{(0)}, \delta \gamma^{(1)}, ..., \delta \gamma^{(n)}, ...$ of standard cosmological perturbation theory. But starting with the correct classical Poisson algebra is relevant for the (future) canonical quantization. Of course, the price to pay is that in our approach $\gamma^{(0)}$ is not a fixed background: it is a {\it dynamical} homogeneous space, its dynamics being generated by the same Hamiltonian that generates the dynamics for the inhomogeneities $\delta \gamma$.}


 {All these differences are unavoidable in our programme. In fact, this feature of treating the homogeneous and inhomogeneous parts on the same footing is precisely what motivates us into thinking that our proposed approach is more suitable for canonical quantization of perturbations \emph{and} background. We conclude in Section \ref{finel} with a discussion on these results.}
\section{The Klein-Gordon-Einstein Theory} \label{clasdeo}
We consider the gravitational field coupled to two Klein-Gordon fields. The system is described by the following action:
\begin{align} \label{action}
S = \int d^4x \sqrt{-g} \left[\frac{1}{2\kappa} R - \frac{1}{2} g^{\mu \nu} \partial_\mu T \partial_\nu T - V_T(T) - \frac{1}{2} g^{\mu \nu} \partial_\mu \phi \partial_\nu \phi - V_\phi(\phi)\right]
\end{align}
where $\kappa = 8\pi G$ and $R$ is the Ricci scalar of the gravitational field $g_{\mu \nu}$. We can distinguish three sectors:
\begin{itemize}
\item the geometric (G) sector, associated to the metric $g_{\mu \nu}$,
\item the time (T) sector, a Klein-Gordon ``clock field'' $T$,
\item the matter (M) sector, a Klein-Gordon field $\phi$.
\end{itemize}
The field $T$ is referred to as time, because the value of its spacially homogenous part (with respect to fixed coordinates), $T^{(0)}$, will be used to parametrize  a one dimensional group of automorphisms acting on the Dirac observables.

To proceed with the canonical quantization of the theory, it is convenient to pass to the Hamiltonian formalism. The usual way to do so is to write the metric $g_{\mu \nu}$ in the ADM form:
\begin{align} \label {ADMmetric}
g_{\mu \nu} dx^\mu dx^\nu = -(N^2 - q_{ab} N^a N^b) dt^2 + 2 q_{ab} N^b dt dx^a + q_{ab} dx^a dx^b
\end{align}
Here, $q_{ab}$ is the spatial metric, i.e. the metric that $g_{\mu \nu}$ induces on the spatial (Cauchy) surface $\Sigma \subset M$. $N$ and $N^a$ are called respectively the \emph{lapse function} and the \emph{shift vector field}, and characterize the spacetime geometry in the directions transversal to  $\Sigma$ embedded  in $M$. In our paper, the latin indices run through the set $\{1, 2, 3\}$ and are characteristic to  the objects living in the {}{tensor bundle} of $\Sigma$. They are raised and lowered by using the spatial metric.
\\
In matrix form, the metric and its inverse are given by
\begin{align} \label{ADMmatrix}
g_{\mu \nu} = \left(
\begin{array}{cccc}
-N^2 + N^a N_a & & N_a & \\
 & & & \\
N_a & & q_{ab} & \\
 & & &
\end{array}
\right),
\ \ \ \ \
g^{\mu \nu} = \left(
\begin{array}{cccc}
-1/N^2 & & N^a/N^2 & \\
 & & & \\
N^a/N^2 & & q^{ab} - N^a N^b/N^2 & \\
 & & & 
\end{array}
\right)
\end{align}
From here, we can perform the canonical analysis of the action (\ref{action}). First, we plug (\ref{ADMmetric}) in (\ref{action}), then use the Codazzi equation to write $R$ in terms of the 3-dimensional Ricci tensor $R^{(3)}$ and of the extrinsic curvature
\begin{align} \label{extcurvature}
K_{ab} & = -\frac{1}{2} (\mathcal{L}_{n} q)_{ab} = -\frac{1}{2N} \left[\dot{q}_{ab} - (\mathcal{L}_{\vec{N}} q)_{ab}\right] = \notag
\\
& = \frac{1}{2N} \left[-\dot{q}_{ab} + \nabla_a N_b + \nabla_b N_a\right]
\end{align}
where $n$ is the unit vector field normal to $\Sigma$.
One obtains
\begin{align} \label{ADMaction1}
S = & \int dt \int d^3x \sqrt{q} \left[\frac{N}{2\kappa} (R^{(3)} + K_{ab} K^{ab} - (q^{ab} K_{ab})^2) + \frac{1}{2N} (\dot{T} - N^a \partial_a T)^2 - \frac{N}{2} q^{ab} \partial_a T \partial_b T - NV_T(T) +\right. \notag
\\
& \left.+ \frac{1}{2N} (\dot{\phi} - N^a \partial_a \phi)^2 - \frac{N}{2} q^{ab} \partial_a \phi \partial_b \phi - NV_\phi(\phi)\right]
\end{align}
Knowing that $R^{(3)}$ involves only  spatial derivatives of $q_{ab}$, it is easy to compute the conjugate momenta:
\begin{align} \label{momenta}
\left\{
\begin{array}{lll}
\pi^{ab} & = \dfrac{\delta S}{\delta \dot{q}_{ab}} & = \dfrac{\sqrt{q}}{2\kappa} (q^{ab} q^{cd} - q^{ac} q^{bd}) K_{cd}
\\
\\
\pi_a & = \dfrac{\delta S}{\delta \dot{N}^{a}} & = 0
\\
\\
\pi & = \dfrac{\delta S}{\delta \dot{N}} & = 0
\\
\\
p_{T} & = \dfrac{\delta S}{\delta \dot{T}} & = \dfrac{\sqrt{q}}{N} (\dot{T} - N^a \partial_a T)
\\
\\
\pi_{\phi} & = \dfrac{\delta S}{\delta \dot{\phi}} & = \dfrac{\sqrt{q}}{N} (\dot{\phi} - N^a \partial_a \phi)
\end{array}
\right.
\end{align}
Using these, we can rewrite (\ref{ADMaction1}) in the canonical form:
\begin{align} \label{ADMaction2}
S = \int dt \int d^3x \left[\pi^{ab} \dot{q}_{ab} + p_{T} \dot{T} + \pi_{\phi} \dot{\phi} - N C - N^a C_a\right]
\end{align}
where
\begin{align} \label{constraints}
\left\{
\begin{array}{lcl}
C & = & \dfrac{2\kappa}{\sqrt{q}} \left[\pi_{ab} \pi^{ab} - \dfrac{1}{2} (q_{ab} \pi^{ab})^2\right] - \dfrac{\sqrt{q}}{2 \kappa} R^{(3)} +
\\
& & + \dfrac{1}{2 \sqrt{q}} p_{T}^2 + \dfrac{\sqrt{q}}{2} q^{ab} \partial_a T \partial_b T + \sqrt{q} V_T(T) + \dfrac{1}{2 \sqrt{q}} \pi_{\phi}^2 + \dfrac{\sqrt{q}}{2} q^{ab} \partial_a \phi \partial_b \phi + \sqrt{q} V_\phi(\phi)
\\
\\
C_a & = & -2 q_{ac} \nabla_b \pi^{bc} + p_T \partial_a T + \pi_\phi \partial_a \phi
\end{array}
\right.
\end{align}
Looking at the action (\ref{ADMaction2}), one can see that the phase  space $\Gamma$ and coordinates thereon in which Poisson brackets have the canonical form are manifest: we may separate them as 
\begin{align} \label{PSsectors1}
\Gamma = \Gamma'_G \times \Gamma_T \times \Gamma_M
\end{align}
where coordinates $(N, N^a, q_{ab}, \pi, \pi_a, \pi^{ab})$ parametrize the geometric sector 
$\Gamma'_G$, coordinates $(T, p_T)$ parametrize the time sector $\Gamma_T$, and $(\phi, \pi_\phi)$ parametrize the matter sector $\Gamma_M$. However, as $N$ and $N^a$ are non-dynamical, we have four primary constraints (per point $x \in \Sigma$): $\pi = 0$ and $\pi_a = 0$. These constraints can be directly solved, and the transformations they generate can be gauge-fixed by choosing arbitrary  $N = N(q_{ab},\pi^{ab},T,p_T,\phi,p_\phi)$ and $N^a = N^a(q_{ab},\pi^{ab},T,p_T,\phi,p_\phi)$: the other variables do not depend on them, so they are all gauge-invariant under this choice. So the phase space of the system reduces to 
\begin{align} \label{PSsectors2}
\Gamma = \Gamma_G \times \Gamma_T \times \Gamma_M
\end{align}
where $\Gamma_G$ is parametrized by $(q_{ab}, \pi^{ab})$ only. The Poisson structure takes the canonical form in those coordinates: in other words, the only nonvanishing brackets are
\begin{align} \label{Falgebra}
\{q_{ab}(x), \pi^{cd}(y)\} = \delta^{(c}_a \delta^{d)}_b \delta^{(3)}(x, y), \ \ \ \ \ \{T(x), p_{T}(y)\} = \delta^{(3)}(x, y), \ \ \ \ \ \{\phi(x), \pi_{\phi}(y)\} = \delta^{(3)}(x, y)
\end{align}
Conservation of the primary constraints under the evolution generated by the Hamiltonian
\begin{align} \label{Hamiltonian}
H = \int d^3x \left[N C + N^a C_a\right]
\end{align}
implies four secondary constraints (per point $x \in \Sigma$): 
\begin{align} \label{theconstraints}
C = 0, \ \ \ \ \ C_a = 0
\end{align} 
It can be shown that these constraints are conserved with respect to $H$, so $(C, C_a)$ constitutes the whole set of constraints. Moreover, their constraint algebra closes, so they form a set of first-class constraints (using Dirac's terminology).

Observables of the theory are those phase space functions $F$ -- called the \emph{Dirac observables} -- that Poisson-commute with all the constraints:
\begin{align} \label{observables}
\{F, C(x)\} = \{F, C_a(x)\} = 0, \ \ \ \mbox{for all} \ x \in \Sigma
\end{align}
This in particular means that any observable $F$ does not evolve, as it commutes with the Hamiltonian (\ref{Hamiltonian}):
\begin{align} \label{evolution}
\frac{d}{dt} F = \{F, H\} = 0
\end{align}
This so-called \emph{problem of time} is solved by introducing a suitable automorphism on the Poisson algebra of Dirac observables. 

It is possible to show that the constraints $(C, C_a)$ encode an important geometrical feature of the theory: diffeomorphism-invariance. Indeed, $C_a$ and $C$ generate the action on the phase space of diffeomorfisms of $\Sigma$ and diffeomorphisms off $\Sigma$ (in the normal direction), respectively. For this reason they are often called respectively \emph{vector constraint} and \emph{scalar constraint}.
\section{Background sector of $\Gamma$} \label{BiG}
We want to consider generic linear perturbations on a fixed background. The ideal background (which is also the physically meaningful one) presents homogeneous and isotropic space $\Sigma$. We are assuming throughout this paper, that $\Sigma$ is a 3-torus endowed with a symmetry group by choosing 6 vector fields: 3 generators of global translations and 3 generators of local rotations. We parametrize the $3$-torus by coordinates $x^1,x^2,x^3\in [0,1)$ (where the interval $[0,1)$ is endowed with the  topology of a circle) --  what we could call the frame of ``generalized cosmological observers''. In terms of these coordinates the symmetry generators are $\partial_a$ and $\epsilon_{abc} x^b \partial^c$. This structure will be used to define the ``background'' sector, a subspace 
\begin{align} \label{thebackground}
\Gamma^{(0)}=\Gamma_G^{(0)}\times \Gamma_T^{(0)}\times\Gamma_M^{(0)}\ \ \subset\ \  \Gamma_G\times \Gamma_T\times\Gamma_M
\end{align}

$\Gamma_G^{(0)}$ is the homogeneous isotropic part of the geometric sector $\Gamma_G$ of the phase space. It consists of points $(q_{ab}^{(0)},\pi^{ab}_{(0)})$ such that the vector fields $\partial_a$ and $\epsilon_{abc} x^b \partial^c$ are their symmetries:
\begin{align} \label{0thGEO}
q_{ab}^{(0)}(x) = e^{2\alpha} \delta_{ab}, \ \ \ \ \ \pi^{ab}_{(0)}(x) = \frac{\pi_{\alpha} e^{-2 \alpha}}{6} \delta^{ab}
\end{align}
where $\alpha$ and $\pi_\alpha$ are constant. Therefore, $\Gamma_G^{(0)}$ is freely parametrized by $(\alpha, \pi_\alpha)$.

In $\Gamma_M^{(0)}$, the matter field $\phi$ is assumed to be absent, in the sense, that 
$\pi_\phi= 0$ and $\phi=\phi_0$, the minimum of the potential $V_\phi$. 
For simplicity let us assume that 
\begin{align} \label{Vmin}
\phi_0 = V_\phi(\phi_0) = 0
\end{align}
although in future it may be also be interesting to study the consequences of the spontanues symmetry breaking in this context. Therefore, $\Gamma_M^{(0)}$ consists of a one point $(0,0)$.    

Contrary to $\phi$, a nontrivial  time field $T$ is necessary. To be consistent with the homogeneity of the space, we choose the background $T$ homogeneous as well: this means that the infinitely many degrees of freedom sitting in $T(x)$ and $p_T(x)$ are reduced to a unique one. Hence,  $(T^{(0)},p_T^{(0)})$ freely parametrize 
$\Gamma_T^{(0)}$ where 
\begin{align} \label{0thTIME}
T(x)\ =\ T^{(0)}, \ \ \ \ \ p_T(x)\ =\ p_T^{(0)}
\end{align}

The subspace $\Gamma^{(0)}$ can be intersected with the constraint surface 
\begin{align} \label{gammasi}
\Gamma_C \subset \Gamma
\end{align}
consisting of solutions to the constraints (\ref{theconstraints}).  
Since everything is spatially homogeneous, the vector constraint $C_a(x)=0$ is automatically satisfied for every point of $\Gamma^{(0)}$. In fact, only the homogeneous part of the scalar constraint survives (below, we denote the restriction of the constraint $C$ to $\Gamma^{(0)}$ by $C^{(0)}$):
\begin{align} \label{homoScal}
C^{(0)}(N) & = \int d^3x N(x) C^{(0)}(x) = e^{-3\alpha} \left[\dfrac{1}{2} (p_T^{(0)})^2 + e^{6\alpha} V_T(T^{(0)}) - \dfrac{\kappa}{12} \pi_{\alpha}^2\right] \int d^3x N(x)
\end{align}
Therefore, at the intersection with the constraint surface $\Gamma_C$, 
the points of $\Gamma^{(0)}$ additionally satisfy a constraint:
\begin{align} \label{theconstraint0}
\dfrac{1}{2} (p_T^{(0)})^2 + e^{6\alpha} V_T(T^{(0)}) - \dfrac{\kappa}{12} \pi_{\alpha}^2\ =\ 0
\end{align}
These points correspond to a FRW spacetime 
\begin{align} \label{FredBobWally}
g^{(0)}_{\mu \nu} dx^\mu dx^\nu \ =\ - dt^2 + e^{2\alpha(t)} \delta_{ab} dx^a dx^b
\end{align}
satisfying Einstein equation with energy-momentum tensor given by the clock field $T=T^{(0)}(t)$ (which satifies Klein-Gordon equation in a spacetime of the form (\ref{FredBobWally})).
The dependence on the variable $t$  of $\alpha,\pi_\alpha, T^{(0)}$ and $p_T^{(0)}$ (upon assumption $N(t)=1$, and recalling that in the chosen coordinates $\int_\Sigma d^3x = 1$) is given by Hamilton equations:
\begin{align} \label{0thHamEq}
\left\{
\begin{array}{lcl}
\dot{\alpha} & = & -\dfrac{\kappa}{6} e^{-3\alpha} \pi_\alpha
\\
\\
\dot{\pi}_\alpha & = & \dfrac{3}{2} e^{-3\alpha} (p_T^{(0)})^2 - \dfrac{\kappa}{4} e^{-3\alpha} \pi_\alpha^2 - 3 e^{3\alpha} V_T
\\
\\
\dot{T}^{(0)} & = & e^{-3\alpha} p_T^{(0)}
\\
\\
\dot{p}_T^{(0)} & = & -e^{3\alpha} \dfrac{\partial V_T}{\partial T}
\end{array}
\right.
\end{align}
A solution of this system of equations which also satisfies the constraint (\ref{theconstraint0}) yields the (dynamical) background FRW metric $g^{(0)}_{\mu \nu}$ on which usual perturbation theory is developed. 

As we are going to see, \emph{in our approach to the full theory we do not intersect the homogeneous isotropic sector $\Gamma^{(0)}$ with the constraint surface $\Gamma_C$, nor we fix the background  dynamics to be of the form} (\ref{0thHamEq}). Instead, all the  homogeneous isotropic sector $\Gamma^{(0)}$ will be coupled to the perturbations and together they will obey the dynamics of the full theory. Not surprisingly, at linear order in the perturbations, the background dynamics does reduce to (\ref{0thHamEq}), so indeed our homogeneous part of the metric $g^{(0)}_{\mu \nu}$ is of the FRW type in that approximation. Nevertheless, it is important to notice that:
\begin{enumerate}
\item[(i)] In our framework -- designed for canonical quantization -- the degrees of freedom $\alpha, \pi_\alpha, T^{(0)}$ and $p_T^{(0)}$ will be treated on the same footing as the remaining degrees of freedom;
\item[(ii)] (\ref{theconstraint0}) and (\ref{0thHamEq}) are true only up to  the linear order: if one considers higher orders, the back reaction is present, and our homogeneous isotropic part will not anymore be a solution to (\ref{theconstraint0})-(\ref{0thHamEq}).
\end{enumerate}
\section{Coordinates on the full phase space} \label{fullcor}
{}{Now, we go back to the full phase space, and define on it some clever coordinate system, adapted to the background-perturbation split (which will be performed in the next section). All formulae appearing in the present section are exact and valid for every point $\gamma \in \Gamma$ of the full phase space.}
\subsection{Extension of the background coordinates to the full phase space $\Gamma$}
The functions parametrizing the homogeneous isotropic subspace $\Gamma^{(0)}$ can be extended to the full phase space.  We do it in the following way. Given $(q_{ab}, \pi^{ab}, T, p_T,\phi,p_\phi)\in \Gamma$, define
\begin{align} \label{backthen}
\left\{
\begin{array}{lcl}
\alpha\ &=&\ \dfrac{1}{2}{\rm ln}\left(\dfrac{1}{3}\delta^{ab}\int_{\Sigma } d^3xq_{ab}\right)
\\
\\
\pi_\alpha\ &=&\ 2e^{2\alpha}\delta_{ab}\int_{\Sigma } d^3x\pi^{ab}
\\
\\
T^{(0)}\ &=&\ \int_\Sigma d^3x T
\\
\\
p_T^{(0)}\ &=&\  \int_\Sigma d^3x p_T
\end{array}
\right.
\end{align}
In this way, $\alpha, \pi_\alpha, T^{(0)}$ and $p_T^{(0)}$ become functions defined on the phase space $\Gamma$ of the full theory. Notice that each of them is defined globally on $\Gamma$.\footnote{
Indeed, $\delta^{ab}q_{ab}(x)> 0$ ensures that even $\alpha$ is defined  globally.
}
We call them \emph{background coordinates} (or background variables) on $\Gamma$.\footnote{
This terminology comes simply from the fact that, in the next section, we will be Taylor-expanding certain phase space functions (namely, the constraints) around a neighbourghood of points $(\alpha, \pi_\alpha, T^{(0)}, 0, ..., 0) \in \Gamma$. As already explained, no dynamical propery is taken into account here.
}
Restricted to the subspace $\Gamma^{(0)}$, they coincide with the coordinates introduced in the previous section and denoted in the same way.
\subsection{Perturbation coordinates on $\Gamma$}
Next,  for every $x\in \Sigma$, on the full phase space $\Gamma$ we define functions $\delta q_{ab}(x), \delta \pi^{ab}(x), \delta T(x), \delta p_T(x), \delta \phi(x), \delta \pi_\phi(x)$ which, together with the background coordinates, form a coordinate system on $\Gamma$:    
\begin{align} \label{perts}
\left\{
\begin{array}{lcl}
\delta q_{ab}(x) & = & q_{ab}(x)\ -\  e^{2\alpha} \delta_{ab}
\\
\\
 \delta\pi^{ab}(x) & = & \pi^{ab}(x)\ -\ \dfrac{\pi_\alpha}{6} e^{-2\alpha} \delta^{ab}
\\
\\
 \delta T(x) & = & T(x)\ -\ T^{(0)}
\\
\\
\delta p_T(x) & = & p_T(x)\ -\ p_T^{(0)}
\\
\\
\delta \phi(x) & = & \phi(x)\ 
\\
\\
\delta\pi_\phi(x) & = & \pi_\phi(x)\ 
\end{array}
\right.
\end{align}
Let us call them \emph{perturbation coordinates}. They satisfy the following identities:
\begin{align} \label{nonHomo}
\int d^3x \delta^{ab} \delta q_{ab}(x) = \int d^3x \delta_{ab} \delta \pi^{ab}(x) = 0, \ \ \ \ \ \int d^3x \delta T(x) = \int d^3x \delta p_T(x) = 0
\end{align}
Such relations constrain the perturbation coordinates.

Using the background coordinates and the perturbation coordinates, we are simply parametrizing points in $\Gamma$ in terms of their suitably defined components in $\Gamma^{(0)} \subset \Gamma$ and the rest. It is good to keep the phase space picture in mind, since later it will be essential that we do not forget about the background sector (as usually done in standard cosmological perturbation theory, by demoting its degrees of freedom to fixed parameters instead of dynamical variables). Indeed, while for the classical theory there is no difference, to properly prepare the set up for quantization (of the full system: perturbations \emph{and} background) this is the correct way to go.
\subsection{Fourier mode-decomposition}
The coordinates $(x^a)=(x^1,x^2,x^3)$ fixed on $\Sigma$ can be also used to introduce a mode-decomposition (or \emph{real Fourier transform}) of fields. First, let us define the usual Fourier transform (with respect to coordinates $(x^{a})$ for $\Sigma$) of a field $f(x)$ by
\begin{align} \label{FTransf}
\tilde{f}(k) = \int d^3x e^{-i k \cdot x} f(x)
\end{align}
We think of $k$ as a spatial vector (i.e., tangent to $\Sigma$), which labels the \emph{mode} of $f$. It takes values in the lattice $\mathcal{L} = (2\pi \mathbb{Z})^3$.

For a real-valued $f$ as our fields, it is $\tilde{f}(-k)\ =\ \overline{\tilde{f}(k)}$. Thus, to isolate the truly independent modes we work with the real Fourier transform. Split the lattice $\mathcal{L}$ into ''positive'', ''negative'' and ''zero'' vectors:
\begin{align} \label{latticeSplit}
\begin{array}{ll}
\mathcal{L}_+ & = \{k \in \mathcal{L} : (k_1 > 0) \lor (k_1 = 0 \land k_2 > 0) \lor (k_1 = k_2 = 0 \land k_3 > 0)\}
\\
\mathcal{L}_- & = \{k \in \mathcal{L} : (k_1 < 0) \lor (k_1 = 0 \land k_2 < 0) \lor (k_1 = k_2 = 0 \land k_3 < 0)\}
\\
\mathcal{L}_0 & = \{0\}
\end{array}
\end{align}
Clearly, we have $\mathcal{L} = \mathcal{L}_+ \cup \mathcal{L}_- \cup \mathcal{L}_0$. Then, we define the real Fourier transform of $f(x)$ as
\begin{align} \label{RFTransf}
\breve{f}(k) = \left\{
\begin{array}{llll}
(\tilde{f}(k) + \tilde{f}(-k))/\sqrt{2} & & & \mbox{if} \ k \in \mathcal{L}_+
\\
(\tilde{f}(k) - \tilde{f}(-k))/i\sqrt{2} & & & \mbox{if} \ k \in \mathcal{L}_-
\\
\tilde{f}(0) & & & \mbox{if} \ k = 0
\end{array}
\right.
\end{align}
More explicitely, we have
\begin{align} \label{RFT}
\breve{f}(k) = \left\{
\begin{array}{llll}
\dfrac{1}{\sqrt{2} } \int d^3x \left(e^{i k \cdot x} + e^{-i k \cdot x}\right) f(x) & & & \mbox{if} \ k \in \mathcal{L}_+
\\
\\
\dfrac{i}{\sqrt{2} } \int d^3x \left(e^{i k \cdot x} - e^{-i k \cdot x}\right) f(x) & & & \mbox{if} \ k \in \mathcal{L}_-
\\
\\
 \int d^3x f(x) & & & \mbox{if} \ k = 0
\end{array}
\right.
\end{align}
\\
Knowing $\breve{f}(k)$, we can easily reconstruct the field: from Fourier antitransform, we write
\begin{align} \label{reconstruction1}
f(x) = \tilde{f}(0) + \sum_{k \in \mathcal{L}_+} e^{ik \cdot x} \tilde{f}(k) + \sum_{k \in \mathcal{L}_-} e^{ik \cdot x} \tilde{f}(k) = \tilde{f}(0) + \sum_{k \in \mathcal{L}_+} \left[e^{ik \cdot x} \tilde{f}(k) + e^{-ik \cdot x} \tilde{f}(-k)\right]
\end{align}
where we used the fact that $k \in \mathcal{L}_-$ is quivalent to $-k \in \mathcal{L}_+$, and then replaced the dummy index $-k$ with $k$. Inverting the definition (\ref{RFTransf}) of $\breve{f}(k)$, one has
\begin{align} \label{reconstruction2}
\sqrt{2} \breve{f}(k) = \tilde{f}(k) + \tilde{f}(-k), \ \ \ \ \ i \sqrt{2} \breve{f}(-k) = \tilde{f}(k) - \tilde{f}(-k), \ \ \ \ \ \mbox{for} \ k \in \mathcal{L}_+
\end{align}
thus, by adding and subtracting these two, one finds respectively
\begin{align} \label{reconstruction3}
\tilde{f}(k) = \frac{1}{\sqrt{2}} \left(\breve{f}(k) + i \breve{f}(-k)\right), \ \ \ \ \ \tilde{f}(-k) = \frac{1}{\sqrt{2}} \left(\breve{f}(k) - i \breve{f}(-k)\right), \ \ \ \ \ \mbox{for} \ k \in \mathcal{L}_+
\end{align}
Replacing these in (\ref{reconstruction1}), we finally obtain the inverse transform:
\begin{align} \label{reconstruction4}
f(x) = \breve{f}(0) + \sum_{k \in \mathcal{L}_+} \left[\breve{f}(k) \frac{e^{ik \cdot x} + e^{-ik \cdot x}}{\sqrt{2}} + i \breve{f}(-k) \frac{e^{ik \cdot x} - e^{-ik \cdot x}}{\sqrt{2}}\right]
\end{align}

Specifically, for our fundamental fields the mode-decomposition is the following:\footnote{
Since we are using the Fourier transform with respect to the fiducial metric, $\delta_{ab}$, the wavevector $k^a$ is \emph{not} the physical momentum. To get the physical momentum, one needs to mutiply by the inverse of the scale factor.
}
\begin{align} \label{pertKexp}
\left\{
\begin{array}{ll}
q_{ab}(x) & = e^{2\alpha} \delta_{ab} + \delta \breve{q}_{ab}(0) + \sum_{k \in \mathcal{L}_+} \left[\delta \breve{q}_{ab}(k) \dfrac{e^{ik \cdot x} + e^{-ik \cdot x}}{\sqrt{2}} + i \delta \breve{q}_{ab}(-k) \dfrac{e^{ik \cdot x} - e^{-ik \cdot x}}{\sqrt{2}}\right]
\\
\\
\pi^{ab}(x) & = \dfrac{\pi_\alpha e^{-2\alpha}}{6} \delta^{ab} + \delta \breve{\pi}^{ab}(0) + \sum_{k \in \mathcal{L}_+} \left[\delta \breve{\pi}^{ab}(k) \dfrac{e^{ik \cdot x} + e^{-ik \cdot x}}{\sqrt{2}} + i \delta \breve{\pi}^{ab}(-k) \dfrac{e^{ik \cdot x} - e^{-ik \cdot x}}{\sqrt{2}}\right]
\\
\\
T(x) & = \breve{T}(0) + \sum_{k \in \mathcal{L}_+} \left[\delta \breve{T}(k) \dfrac{e^{ik \cdot x} + e^{-ik \cdot x}}{\sqrt{2}} + i \delta \breve{T}(-k) \dfrac{e^{ik \cdot x} - e^{-ik \cdot x}}{\sqrt{2}}\right]
\\
\\
p_T(x) & = \breve{p}_T(0) + \sum_{k \in \mathcal{L}_+} \left[\delta \breve{p}_T(k) \dfrac{e^{ik \cdot x} + e^{-ik \cdot x}}{\sqrt{2}} + i \delta \breve{p}_T(-k) \dfrac{e^{ik \cdot x} - e^{-ik \cdot x}}{\sqrt{2}}\right]
\\
\\
\phi(x) & = \delta \breve{\phi}(0) + \sum_{k \in \mathcal{L}_+} \left[\delta \breve{\phi}(k) \dfrac{e^{ik \cdot x} + e^{-ik \cdot x}}{\sqrt{2}} + i \delta \breve{\phi}(-k) \dfrac{e^{ik \cdot x} - e^{-ik \cdot x}}{\sqrt{2}}\right]
\\
\\
\pi_\phi(x) & = \delta \breve{\pi}_\phi(0) + \sum_{k \in \mathcal{L}_+} \left[\delta \breve{\pi}_\phi(k) \dfrac{e^{ik \cdot x} + e^{-ik \cdot x}}{\sqrt{2}} + i \delta \breve{\pi}_\phi(-k) \dfrac{e^{ik \cdot x} - e^{-ik \cdot x}}{\sqrt{2}}\right]
\end{array}
\right.
\end{align}
The $k = 0$ mode corresponds to the homogeneous part. So we have
\begin{align} \label{nullKpart}
\breve{T}(0) = T^{(0)}, \ \ \ \ \ \breve{p}_T(0) = p_{T}^{(0)}
\end{align}
For the metric perturbations, the $k = 0$ case is nonzero only as a traceless matrix:
\begin{align} \label{pertTraceless}
\delta^{ab} \delta \breve{q}_{ab}(0) = 0, \ \ \ \ \ \delta_{ab} \delta \breve{\pi}^{ab}(0) = 0
\end{align}
As a confirmation of this, notice that the $k = 0$ mode is by definition $\breve{f}(0) = \int d^3x f(x)$. But since (\ref{nonHomo}) holds, we have directly the constraints just mentioned.
\subsection{The scalar, vector and tensor modes of the metric}
The treatment of the metric and its momentum requires some more work. Indeed, since $\delta \breve{q}_{ab}(k)$ defines a symmetric $3 \times 3$ matrix per each $k$, once we fix the mode $k$ we can expand $\delta \breve{q}_{ab}(k)$ on a basis for the $6$-dimensional vector space of simmetric $3 \times 3$ matrices. A good basis in the space of symmetric bi-covariant tensors is $\{A^m_{ab}\}$ ($m = 1, ..., 6$) defined by
\begin{align} \label{basisSymmatrices}
\begin{array}{ll}
A^1_{ab} = \delta_{ab}, & A^2_{ab} = \dfrac{k_a k_b}{k^2} - \dfrac{1}{3} \delta_{ab}
\\
\\
A^3_{ab} = \dfrac{1}{\sqrt{2}} (k_a v_b + k_b v_a), & A^4_{ab} = \dfrac{1}{\sqrt{2}} (k_a w_b + k_b w_a)
\\
\\
A^5_{ab} = \dfrac{k^2}{\sqrt{2}} (v_a w_b + v_b w_a), & A^6_{ab} = \dfrac{k^2}{\sqrt{2}} (v_a v_b - w_a w_b)
\end{array}
\end{align}
where $v$ and $w$ are spatial vectors forming with $k$ an orthogonal basis of the momentum space $\mathbb{R}^3$ (with respect to the fiducial metric $\delta_{ab}$, which is also used to raise and lower indices for $v$, $w$ and $k$). The normalization of $v$ and $k$ is chosen to be $v^2 = w^2 = 1/k^2$.

The subspaces spanned respectively by $(A^1, A^2)$, $(A^3, A^4)$ and $(A^5, A^6)$ are said to comprise the \emph{scalar modes}, the \emph{vector modes} and the \emph{tensor modes}. They have the following properties:
\begin{itemize}
\item tensor matrices satisfy $k^a A^m_{ab}(k) = 0$
\item vector matrices satisfy $k^a k^b A^m_{ab}(k) = 0$
\end{itemize}
Also, notice that all matrices except from $A^1_{ab}$ satisfy
\begin{align} \label{deltaProperty}
\delta^{ab} A^m_{ab}(k) = 0
\end{align}
We decompose $\delta \breve{q}_{ab}(k)$ in this basis, 
\begin{align} \label{qmDEF}
\delta \breve{q}_{ab}(k) = q_m(k) A^m_{ab}(k)
\end{align}
where $q_m(k)$ denotes the $m$th component, for $m=1,...,6$.

Similarly, one can expand $\delta \breve{\pi}^{ab}(k)$ on the dual basis $\{A_m^{ab}\}$:
\begin{align} \label{pmDEF}
\delta \breve{\pi}^{ab}(k) = p^m(k) A_m^{ab}(k)
\end{align}
The dual basis is given by
\begin{align} \label{dualSymmatrices}
\begin{array}{ll}
A_1^{ab} = \dfrac{1}{3} \delta^{ab}, & A_2^{ab} = \dfrac{3}{2} \left(\dfrac{k^a k^b}{k^2} - \dfrac{1}{3} \delta^{ab}\right)
\\
\\
A_3^{ab} = \dfrac{1}{\sqrt{2}} (k^a v^b + k^b v^a), & A_4^{ab} = \dfrac{1}{\sqrt{2}} (k^a w^b + k^b w^a)
\\
\\
A_5^{ab} = \dfrac{k^2}{\sqrt{2}} (v^a w^b + v^b w^a), & A_6^{ab} = \dfrac{k^2}{\sqrt{2}} (v^a v^b - w^a w^b)
\end{array}
\end{align}
It is easy to check the duality, i.e. that $Tr(A_m A^n) = A_m^{ab} A^n_{ba} = A_m^{ab} A^n_{ab} = \delta^{n}_{m}$. Moreover, all these $A$ matrices are normalized with respect to the scalar product induced by the fiducial meric $\delta_{ab}$, namely
\begin{align} \label{scalprodDEF}
(A, A') = \delta^{ac} \delta^{bd} A_{ab} A'_{cd}, \ \ \ \ \ (A, A') = \delta_{ac} \delta_{bd} A^{ab} A'^{cd}
\end{align}
except for $A^1, A^2, A_1, A_2$, for which we have
\begin{align} \label{scalprodscal}
(A^1, A^1) = 3, \ \ \ \ \ (A^2, A^2) = 2/3, \ \ \ \ \ (A_1, A_1) = 1/3, \ \ \ \ \ (A_2, A_2) = 3/2
\end{align}
These scalar matrices are left non-normalized to keep the agreement with formule in \cite{Lang}. Note however that in our case the matrices $A$ do not involve any dynamical variable (in particular, they do not depend on $\alpha$), so no nontrivial Poisson algebra is hidden in expansions (\ref{qmDEF}) and (\ref{pmDEF}): everything is contained in our new variables $q_m(k)$ and $p^m(k)$. Finally, note that it is always possible to choose $v$ and $w$ in such a way that $A$ are symmetric under $k \rightarrow -k$: we do this, so in the following we will have
\begin{align} \label{symmetryA}
A^m_{ab}(k) = A^m_{ab}(-k), \ \ \ \ \ A_m^{ab}(k) = A_m^{ab}(-k)
\end{align}
\subsection{Summary of the section: resulting coordinates on $\Gamma$}
Let us summarize. We have introduced on the full phase space $\Gamma$ the following system of coordinates:
\begin{itemize}
\item the background coordinates: 4 numbers $(\alpha, \pi_\alpha, T^{(0)}, p_T^{(0)})$,
\item the perturbation coordinates (homogeneous part, that is, for $k = 0$): 12 numbers $(\delta \breve{q}_{ab}(0), \delta \breve{\pi}^{ab}(0), \delta \breve{\phi}(0), \delta \breve{\pi}_\phi(0))$,
\item the perturbation coordinates (inhomogeneous part, that is, for $k \in \mathcal{L} - \{0\}$): 16 numbers per each $k$, $(q_m(k), p^m(k), \delta \breve{T}(k), \delta \breve{p}_T(k), \delta \breve{\phi}(k), \delta \breve{\pi}_\phi(k))$.
\end{itemize}
Again, we remark that this terminology is in function of the next section (and the remainder of the paper), but up to now the ''perturbation components'' are defined for every point in phase space $\Gamma$, and are finite.

The fundamental Poisson algebra of the original ADM variables straightforwardly induces the following Poisson algebra on the new ones:
\begin{align} \label{PoissonPerts}
\begin{array}{c}
\{\alpha, \pi_{\alpha}\} = 1, \ \ \ \ \ \{T^{(0)}, p_T^{(0)}\} = 1, \ \ \ \ \ \{\delta \breve{q}_{ab}(0), \delta \breve{\pi}^{cd}(0)\} = \delta^c_{(a} \delta^d_{b)} - \frac{1}{3}\delta^{cd}\delta_{ab}, \ \ \ \ \ \{\delta \breve{\phi}(0), \delta \breve{\pi}_\phi(0)\} = 1
\\
\\
\{q_{m}(k), p^n(k')\} = \delta^n_m \delta_{k, k'}, \ \ \ \ \ \{\delta \breve{T}(k), \delta \breve{p}_T(k')\} = \delta_{k, k'}, \ \ \ \ \ \{\delta \breve{\phi}(k), \delta \breve{\pi}_\phi(k')\} = \delta_{k, k'}
\end{array}
\end{align}
for $k, k' \neq 0$, and the remaining Poisson brackets vanish.

 {This is a good point to compare our approach with the standard one. In the standard perturbation theory one would fix a specific background initial data $\tilde{\gamma}^{(0)}$ admitting a symmetry group isomorphic to the symmetry group of the flat $3$-torus. What we did, instead, is we fixed the $3$-torus and the symmetry group (by choosing 6 vector fields) with no reference to any specific point in the phase space. Next, given any point  $\gamma$ (in general with no symmetry vector fields), we used it to define new initial data $\gamma^{(0)}$ via the integrals in (\ref{backthen}).
The new data is symmetric (homogeneous and isotropic). Having such homogeneous isotropic part $\gamma^{(0)}$, we defined the remaining part $\delta \gamma$. This decomposion is unique for each point $\gamma$ of the phase space, given the symmetry group. Now, contrary to standard cosmological perturbation theory, we are going to write an expansion of, say, energy density $\rho$ as
\begin{align}
\rho(\gamma) = \rho^{(0)}(\gamma^{(0)}) + \rho^{(1)}(\gamma^{(0)}, \delta \gamma) + \rho^{(2)}(\gamma^{(0)}, \delta \gamma) + ...
\end{align}
where $\rho^{(0)}$ is not only the energy density of a background, but is simply the part of the energy density $\rho(\gamma)$ which is function only of the homogeneous isotropic part. Similarly, $\rho^{(1)}$ is the part linear in $\delta \gamma$, $\rho^{(2)}$ is the part quadratic in $\delta \gamma$, etc.\footnote{
 {To see this explicitely, we refer the reader to Section \ref{constup}, where we will be expanding in this way the constraints.}
}
Our expansion is unique given the background symmetry group, pretty much as the standard cosmological perturbation expansion is unique given the symmetric background spacetime.
}
\section{The constraints up to the first order} \label{constup}
\subsection{The expansion}
We now turn to the constraints of the theory. At this point, it is convenient to expand them for ``small'' perturbation variables (\ref{perts}). That is, we are applying the Taylor expansion formally given by
\begin{align} \label{taylor}
F(f^{(0)} + \delta f) = F(f^{(0)}) + F'(f^{(0)}) \delta f + \dfrac{1}{2} F''(f^{(0)}) \delta f^2 + O(\delta f^3)
\end{align}
In our case, $f$ stands for the fields, $f^{(0)}$ for the background variables, and $\delta f$ for the perturbation variables. The
decomposition $f=f^{(0)} + \delta f$ reads
\begin{align} \label{perts2}
\left\{
\begin{array}{ll}
q_{ab}(x) & =    e^{2\alpha} \delta_{ab}\ +\ \delta q_{ab}(x)
\\
\\
\pi^{ab}(x) & =   \dfrac{\pi_\alpha}{6} e^{-2\alpha} \delta^{ab}\ +\ \delta\pi^{ab}(x)
\\
\\
T(x) & =   T^{(0)}\ +\ \delta T(x)\ 
\\
\\
p_T(x) & =    p_T^{(0)}\ +\ \delta p_T(x)
\\
\\
\phi(x) & =  \delta\phi(x)\ 
\\
\\
\pi_\phi(x) & =   \delta\pi_\phi(x)
\end{array}
\right.
\end{align}
In light of the decomposition (\ref{perts2}), we are to expand accordingly the scalar and vector constraints:
\begin{align} \label{constrexpansion}
C(x) = C^{(0)}(x) + C^{(1)}(x) + C^{(2)}(x) + O(\delta f^3), \ \ \ \ \ C_a(x) = C^{(0)}_a(x) + C^{(1)}_a(x) + C^{(2)}_a(x) + O(\delta f^3)
\end{align}
where $C^{(0)}$ (and $C^{(0)}_a$) collects all the terms which are 0th order in the perturbation variables $\delta f$ (i.e., the first term in the Taylor expansion), $C^{(1)}$ (and $C^{(1)}_a$) collects the 1st order terms (the second term in the Taylor expansion), and $C^{(2)}$ (and $C^{(2)}_a$) collects the 2nd order terms (third term in the Taylor expansion). The idea is thus that we retain  the \emph{full} constraints and deal with their exact solutions, but we write such solutions explicitly only up to the first order in the perturbation variables.
More precisely, we have that the first order expansion of the solutions satisfies
\begin{align} \label{higherOrds1}
C^{(0)}(x) + C^{(1)}(x) = O(\delta f^2), \ \ \ \ \ C^{(0)}_a(x) + C^{(1)}_a(x) = O(\delta f^2)
\end{align}
Therefore, in this linear approximation we will solve
\begin{align} \label{higherOrds2}
C^{(0)}(x) + C^{(1)}(x) = 0, \ \ \ \ \ C^{(0)}_a(x) + C^{(1)}_a(x) = 0
\end{align}
and ignore  $C^{(n)}$ and $C^{(n)}_a$ for $n=2,3,...$. On the other hand, we will  retain $C^{(2)}$ to describe the dynamics at first order.

Now, using (\ref{RFT}), we write the real Fourier transforms of the constraints:
\begin{align} \label{RFTconstraints}
\left\{
\begin{array}{llllll}
\breve{C}(0) & = &  C^{(0)} \int d^3x + \int d^3x C^{(1)}(x) + \int d^3x C^{(2)}(x) \ +\ O(\delta f^3)& & &
\\
\\
\breve{C}(k) & = & \dfrac{1}{\sqrt{2}} \int d^3x \left(e^{ik \cdot x} + e^{-ik \cdot x}\right) \left[C^{(0)} + C^{(1)}(x) + C^{(2)}(x)\right] \ +\ O(\delta f^3)& & & \mbox{if} \ k \in \mathcal{L}_+
\\
\\
\breve{C}(k) & = & \dfrac{i}{\sqrt{2}} \int d^3x \left(e^{ik \cdot x} - e^{-ik \cdot x}\right) \left[C^{(0)} + C^{(1)}(x) + C^{(2)}(x)\right] \ +\ O(\delta f^3)& & & \mbox{if} \ k \in \mathcal{L}_-
\\
\\
\breve{C}_a(0) & = &  C_a^{(0)} \int d^3x + \int d^3x C_a^{(1)}(x) + \int d^3x C_a^{(2)}(x) \ +\ O(\delta f^3)& & &
\\
\\
\breve{C}_a(k) & = & \dfrac{1}{\sqrt{2}} \int d^3x \left(e^{ik \cdot x} + e^{-ik \cdot x}\right) \left[C_a^{(0)} + C_a^{(1)}(x) + C_a^{(2)}(x)\right] \ +\ O(\delta f^3) & & & \mbox{if} \ k \in \mathcal{L}_+
\\
\\
\breve{C}_a(k) & = & \dfrac{i}{\sqrt{2}} \int d^3x \left(e^{ik \cdot x} - e^{-ik \cdot x}\right) \left[C_a^{(0)} + C_a^{(1)}(x) + C_a^{(2)}(x)\right] \ +\ O(\delta f^3) & & & \mbox{if} \ k \in \mathcal{L}_-
\end{array}
\right.
\end{align}

These contain many terms, but the following remarks will help us simplifying them:
\begin{itemize}
\item As already pointed out above, $C^{(0)}_a$ vanishes identically.
\item We already said that we disregard the second order terms \emph{except} for the dynamics. Since as far as $H$ is concerned we are free to choose the lapse and the shift, we will select $N(x) = 1$ and $N^a(x) = 0$ in (\ref{Hamiltonian}), thereby obtaining simply $H = \int d^3x C(x) = \breve{C}(0)$. It follows that we must retain the second order only in $\breve{C}(0)$.
\item The terms $\int d^3x C^{(1)}(x)$ and $\int d^3x C^{(1)}_a(x)$ can be seen to vanish because of (\ref{nonHomo}). Also, one sees that $C^{(0)} \int d^3x (e^{i k \cdot x} \pm e^{-i k \cdot x}) \sim C^{(0)} \delta_{k, 0}$, so it vanishes for all $k \neq 0$.
\end{itemize}
Applying these remarks, we see that:
\begin{itemize}
\item[(i)] $\breve{C}_{a}(0) = O(\delta f^2)$ is identically satisfied
\item[(ii)] using (\ref{homoScal}), the constraint $\breve C(0)$ reads
\begin{equation} \label{C(k=0)}
\breve C(0)\ =\  e^{-3\alpha} \left[\dfrac{1}{2} (p_T^{(0)})^2 + e^{6\alpha} V_T(T^{(0)}) - \dfrac{\kappa}{12} \pi_{\alpha}^2\right]\ +O(\delta f^2), 
\end{equation}
{}{which will later be used to solve for $p_T^{(0)}$ as a function of the other background variables.}
\item[(iii)] we are left with 4 constraints per each $k \neq 0$:
\begin{align} \label{RFTremaining}
\left\{
\begin{array}{llllll}
\breve{C}(k) & = & \dfrac{1}{\sqrt{2}} \int d^3x \left(e^{ik \cdot x} + e^{-ik \cdot x}\right) C^{(1)}(x) \ +\ O(\delta f^2) & & & \mbox{if} \ k \in \mathcal{L}_+
\\
\\
\breve{C}(k) & = & \dfrac{i}{\sqrt{2}} \int d^3x \left(e^{ik \cdot x} - e^{-ik \cdot x}\right) C^{(1)}(x) +\ O(\delta f^2) & & & \mbox{if} \ k \in \mathcal{L}_-
\\
\\
\breve{C}_a(k) & = & \dfrac{1}{\sqrt{2}} \int d^3x \left(e^{ik \cdot x} + e^{-ik \cdot x}\right) C_a^{(1)}(x) +\ O(\delta f^2) & & & \mbox{if} \ k \in \mathcal{L}_+
\\
\\
\breve{C}_a(k) & = & \dfrac{i}{\sqrt{2}} \int d^3x \left(e^{ik \cdot x} - e^{-ik \cdot x}\right) C_a^{(1)}(x) +\ O(\delta f^2)& & & \mbox{if} \ k \in \mathcal{L}_-
\end{array}
\right.
\end{align}
\end{itemize}
\subsection{Explicit form}
In order to find their explicit form, we need to first compute the linearized constaints as function of a space point $x$, $C^{(1)}(x)$ and $C^{(1)}_a(x)$. Plugging the decompositions (\ref{perts2}) into (\ref{constraints}) and keeping only the terms linear in the perturbation variables $\delta f$, we find\footnote{
We dropped all the terms coming from the matter field $\phi$, since they always involve a 0th order factor, i.e. either $\phi^{(0)}$ or $\pi_\phi^{(0)}$, which are zero by definition. In this sense, the role of $\phi$ of a ''test field'' is mathematically justified.
}
\begin{align} \label{1stConstraints}
\left\{
\begin{array}{ll}
C^{(1)} = & e^{-3\alpha} \left[\dfrac{e^{6\alpha}}{2\kappa} \left(q^{ab}_{(0)} q^{cd}_{(0)} - q^{ac}_{(0)} q^{bd}_{(0)}\right) \partial_a \partial_b \delta q_{cd} - \dfrac{1}{4} \left(\dfrac{\kappa \pi_\alpha^2}{18} + (p_T^{(0)})^2\right) q^{ab}_{(0)} \delta q_{ab} - \dfrac{\kappa \pi_\alpha}{3} q^{(0)}_{ab} \delta \pi^{ab} +\right.
\\
& \left.+ p_T^{(0)} \delta p_T + \dfrac{e^{6\alpha}}{2} V_T(T^{(0)}) q_{(0)}^{ab} \delta q_{ab} + e^{6\alpha} V_T'(T^{(0)}) \delta T\right]
\\
\\
C^{(1)}_a = & \pi_{(0)}^{bc} \partial_a \delta q_{bc} - 2 q^{(0)}_{ab} \partial_c \delta \pi^{bc} - 2 \pi_{(0)}^{bc} \partial_c \delta q_{ab} + p_T^{(0)} \partial_a \delta T
\end{array}
\right.
\end{align}
In obtaining these linearized constraints, we used some nontrivial facts:
\begin{itemize}
\item The determinant $q$ gets contributions only from the diagonal terms (since at 0th order the non-diagonal terms are zero), so one has
\begin{align} \label{prop:determinant}
q = q^{(0)} + \delta q_{11} q^{(0)}_{22} q^{(0)}_{33} + q^{(0)}_{11} \delta q_{22} q^{(0)}_{33} + q^{(0)}_{11} q^{(0)}_{22} \delta q_{33} = e^{6 \alpha} + e^{4 \alpha} \delta^{ab} \delta q_{ab}
\end{align}
\item For the spatial Ricci scalar $R^{(3)}$ we used the relation
\begin{align} \label{prop:curvature1}
R^{(3)} = q^{ab} \left(\partial_c \Gamma^{\ c}_{a \ b} - \partial_b \Gamma^{\ c}_{a \ c} + \Gamma^{\ d}_{a \ b} \Gamma^{\ c}_{c \ d} - \Gamma^{\ d}_{a \ c} \Gamma^{\ c}_{b \ d}\right)
\end{align}
However, the last two terms are second order in $\delta f$ because they involve products of two 1st order objects (namely, the spatial derivatives of $q_{ab}$ sitting in Christoffel symbols, which are 1st order because the 0th order metric is homogeneous). So one is left with 
\begin{align} \label{prop:curvature2}
R^{(3)} = \left(q_{(0)}^{ac} q_{(0)}^{bd} - q_{(0)}^{ab} q_{(0)}^{cd}\right) \partial_a \partial_b \delta q_{cd} \ +\ O(\delta f^2)
\end{align}
\item Since $\pi^{ab}$ is a tensor density of weight $1$, its covariant derivative is
\begin{align} \label{prop:momentumdensity}
\nabla_a \pi^{bc} = \partial_a \pi^{bc} + \Gamma^{\ b}_{a \ d} \pi^{dc} + \Gamma^{\ c}_{a \ d} \pi^{bd} - \Gamma^{\ d}_{d \ a} \pi^{bc}
\end{align}
Notice that in $\nabla_b \pi^{bc}$ which appears in the  vector constraint, the last term  cancels with the second one.
\end{itemize}

At this point, we can compute the real Fourier transforms (\ref{RFTremaining}). To see how it works, consider the first one, $\breve{C}(k)$ for $k \in \mathcal{L}_+$. It is:
\begin{align} \label{RFconstrcomp1}
E(k)\ :=\ & \breve{C}^{(1)}(k) = \dfrac{1}{\sqrt{2}} \int d^3x \left(e^{ik \cdot x} + e^{-ik \cdot x}\right) C^{(1)}(x) = \notag
\\
=\ & \dfrac{e^{-3\alpha}}{\sqrt{2}} \int d^3x \left(e^{ik \cdot x} + e^{-ik \cdot x}\right) \left[\dfrac{e^{6\alpha}}{2\kappa} \left(q^{ab}_{(0)} q^{cd}_{(0)} - q^{ac}_{(0)} q^{bd}_{(0)}\right) \partial_a \partial_b \delta q_{cd} - \dfrac{1}{4} \left(\dfrac{\kappa \pi_\alpha^2}{18} + (p_T^{(0)})^2\right) q^{ab}_{(0)} \delta q_{ab} -\right. \notag
\\
& \left.- \dfrac{\kappa \pi_\alpha}{3} q^{(0)}_{ab} \delta \pi^{ab} + p_T^{(0)} \delta p_T + \dfrac{e^{6\alpha}}{2} V_T(T^{(0)}) q_{(0)}^{ab} \delta q_{ab} + e^{6\alpha} V_T'(T^{(0)}) \delta T\right] = \notag
\\
=\ & \dfrac{e^{-3\alpha}}{\sqrt{2}} \int d^3x \left(e^{ik \cdot x} + e^{-ik \cdot x}\right) \left[-\dfrac{e^{6\alpha}}{2\kappa} \left(q^{ab}_{(0)} q^{cd}_{(0)} - q^{ac}_{(0)} q^{bd}_{(0)}\right) k_a k_b \delta q_{cd} - \dfrac{1}{4} \left(\dfrac{\kappa \pi_\alpha^2}{18} + (p_T^{(0)})^2\right) q^{ab}_{(0)} \delta q_{ab} -\right. \notag
\\
& \left.- \dfrac{\kappa \pi_\alpha}{3} q^{(0)}_{ab} \delta \pi^{ab} + p_T^{(0)} \delta p_T + \dfrac{e^{6\alpha}}{2} V_T(T^{(0)}) q_{(0)}^{ab} \delta q_{ab} + e^{6\alpha} V_T'(T^{(0)}) \delta T\right] = \notag
\\
=\ & e^{-3\alpha} \left[-\dfrac{e^{6\alpha}}{2\kappa} \left(q^{ab}_{(0)} q^{cd}_{(0)} - q^{ac}_{(0)} q^{bd}_{(0)}\right) k_a k_b \delta \breve{q}_{cd} - \dfrac{1}{4} \left(\dfrac{\kappa \pi_\alpha^2}{18} + (p_T^{(0)})^2\right) q^{ab}_{(0)} \delta \breve{q}_{ab} -\right. \notag
\\
& \left.- \dfrac{\kappa \pi_\alpha}{3} q^{(0)}_{ab} \delta \breve{\pi}^{ab} + p_T^{(0)} \delta \breve{p}_T + \dfrac{e^{6\alpha}}{2} V_T(T^{(0)}) q_{(0)}^{ab} \delta \breve{q}_{ab} + e^{6\alpha} V_T'(T^{(0)}) \delta \breve{T}\right]
\end{align}
where in the last step we used (\ref{RFT}) with respect to the perturbation variables. We can rewrite this as
\begin{align} \label{crazyScalC2}
E(k) = & - \dfrac{e^{-\alpha}}{2\kappa} \left(k^2 \delta^{ab} - k^a k^b\right) A^m_{ab}(k) q_m(k) - \dfrac{e^{-5\alpha}}{4} \left(\dfrac{\kappa \pi_\alpha^2}{18} + (p_T^{(0)})^2 - 2e^{6\alpha} V_T(T^{(0)})\right) \delta^{ab} A^m_{ab}(k) q_m(k) - \notag
\\
& - \dfrac{\kappa \pi_\alpha e^{-\alpha}}{3} \delta_{ab} A_m^{ab}(k) p^m(k) + e^{-3\alpha} p_T^{(0)} \delta \breve{p}_T(k) + e^{3\alpha} V_T'(T^{(0)}) \delta \breve{T}(k)
\end{align}
having replaced the explicit expression for the background variables, and having expanded the perturbation variables of the metric (and conjugate momentum) in the $\{A_{ab}^m\}$ basis. Repeating the computation for $\breve{C}(k)$ in the $k \in \mathcal{L}_-$ case, we obtain the same object, which is then regarded as the scalar constraint satisfied by each mode $k \in \mathcal{L} - \{0\}$:
\begin{align} \label{kScalarC}
E(k) = & - \dfrac{3 e^{-5\alpha}}{4} \left(\dfrac{\kappa \pi_\alpha^2}{18} + (p_T^{(0)})^2 - 2e^{6\alpha} V_T(T^{(0)})\right) q_1(k) - \dfrac{e^{-\alpha}}{\kappa} k^2 q_1(k) + \dfrac{e^{-\alpha}}{3 \kappa} k^2 q_2(k) - \notag
\\
& - \dfrac{\kappa \pi_\alpha e^{-\alpha}}{3} p^1(k) + e^{-3\alpha} p_T^{(0)} \delta \breve{p}_T(k) + e^{3\alpha} V_T'(T^{(0)}) \delta \breve{T}(k)
\end{align}
where we used (\ref{basisSymmatrices}) and (\ref{dualSymmatrices}) to write $E(k)$ as a function of the dynamical variables only.

One obtains $\breve{C}_a^{(1)}(k)$ in a similar way. Moreover, since $\breve{C}_a(k)$ really encodes 3 constraints, we need to separate them. To do this, we project $\breve{C}_a(k)$ along the 3 orthogonal vectors $k$, $v$ and $w$ at our disposal:
\begin{align} \label{crazyVectC1}
M(k) = k^a \breve{C}_a(k), \ \ \ \ \ V(k) = v^a \breve{C}_a(k), \ \ \ \ \ W(k) = w^a \breve{C}_a(k)
\end{align}
Explicitely, these 3 constraints satisfied by each mode $k \in \mathcal{L} - \{0\}$ are given by
\begin{align} \label{kVectorC}
\left\{
\begin{array}{ll}
M(k) & = \dfrac{\pi_\alpha e^{-2\alpha}}{6} q_1(k) - \dfrac{2 \pi_\alpha e^{-2\alpha}}{9} q_2(k) - \dfrac{2 e^{2\alpha}}{3} p^1(k) - 2 e^{2\alpha} p^2(k) + p_T^{(0)} \delta \breve{T}(k)
\\
\\
V(k) & = \dfrac{\pi_\alpha e^{-2\alpha}}{3} q_3(k) + 2 e^{2\alpha} p^3(k)
\\
\\
W(k) & = \dfrac{\pi_\alpha e^{-2\alpha}}{3} q_4(k) + 2 e^{2\alpha} p^4(k)
\end{array}
\right.
\end{align}

Notice that, apart from the background variables, the linearized constraints $E(k)$ and $M(k)$ only involve the scalar modes (namely $q_1(k)$, $q_2(k)$, $p^1(k)$, $p^2(k)$, $\delta \breve{T}(k)$ and $\delta \breve{p}_T(k)$), whereas $V(k)$ and $W(k)$ only involve vector modes (namely $q_3(k)$, $q_4(k)$, $p^3(k)$ and $p^4(k)$). It follows that the linearized constraints  $E$ and $M$ only constrain the scalar sector, while $V$ and $W$ constrain the vector sector. Interestingly, the tensor sector is left completely unconstrained.
\subsection{Preliminary analysis of independent degrees of freedom}
Recalling that each constraint reduces the number of degrees of freedom by $2$ ($1$ for the reduction onto the constraint surface, and $1$ for fixing a gauge -- or equivalently for identifying  each $1$-dimensional orbit  with a single point), we can then proceed with the counting of degrees of freedom:
\begin{itemize}

\item For $k=0$, up to the first order the constraints  $\breve C(0)$ and $\breve C_a(0)$  constrain only the background coordinates $\alpha, \pi_\alpha, T^{(0)}$ and $p_T^{(0)}$ by (\ref{C(k=0)}). That constraint can be solved for $p_T^{(0)}$ and used to gauge fix $T^{(0)}$. On the other hand the traceless variables $\delta \breve q_{ab}(0)$,  $\delta \breve \pi^{ab}(0)$, and the variables $\breve \phi(0)$,  $\breve\pi_\phi(0)$ are unconstrained. 
 
\item For every $k\not= 0$ and given background coordinates $\alpha, \pi_\alpha, T^{(0)}$ and $p_T^{(0)}$, the scalar sector of the phase space is coordinatized by $(q_1(k), p^1(k), q_2(k), p^2(k), \delta \breve{T}(k), \delta \breve{p}_T(k), \delta \breve{\phi}(k), \delta \breve{\pi}_\phi(k))$, so it has dimension $8$. On it there are the $2$ constraints $E(k)$ and $M(k)$, so the dimension is reduced by $2 \cdot 2 = 4$. We conclude that the corresponding sector of the reduced phase space has dimension $8 - 4 = 4$: in other words, there are $4$ gauge-invariant (i.e., physical) scalar degrees of freedom. 2 of them can be chosen to be $(\delta \breve{\phi}(k), \delta \breve{\pi}_\phi(k))$. The other 2 independent degrees of freedom  can be chosen to be particular functions of the remaining variables $(q_1(k), p^1(k), q_2(k), p^2(k), \delta \breve{T}(k), \delta \breve{p}_T(k))$ (see later).

\item For every $k\not= 0$ and given $\alpha, \pi_\alpha$, the vector sector of the phase space is coordinatized by $(q_3(k), p^3(k), q_4(k), p^4(k))$, so it has dimension $4$. Imposed on those variables there are $2$ constraints, $V(k)$ and $W(k)$, so the dimension is reduced by $2 \cdot 2 = 4$. We conclude that the reduced phase space has dimension $4 - 4 = 0$: in other words, vector modes are completely non-physical, and can be gauged away.

\item For every $k\not= 0$ and given background coordinates $\alpha, \pi_\alpha, T^{(0)}$ and $p_T^{(0)}$,  the tensor sector of phase space is coordinatized by $(q_5(k), p^5(k), q_6(k), p^6(k))$, so it has dimension $4$. Up to the first order, there are no constraints imposed on those variables, so there is no reduction in dimension: we conclude that the reduced phase space has dimension $4$, i.e. there are $4$  degrees of freedom. Obviously, they are $(q_5(k), p^5(k), q_6(k), p^6(k))$ themselves. These (or rather, the $2$ configuration variables $q_5$ and $q^6$) are the $2$ polarizations of the graviton. 
\end{itemize}
For the reader familar with the  perturbative approach to the canonical gravity 
often used in cosmology, it is important to notice the difference of our approach. In the standard approach, one often introduces functions of the perturbation coordinates which Poisson-commute with the linearised constraints  $E$, $M$, $V$, and $W$. In that approach they are Dirac observables and play a fundamental role.
In our approach, the transformations generated by the 2nd order terms of the Taylor expansion of the full constraints also contribute to the gauge transformations.
Moreover, for us also $\breve C^{(2)}(0)$ is a generator of gauge transformations. 
Therefore, Poisson commuting with $E$, $M$, $V$, and $W$ is not sufficient to be gauge invariant in the sense of the current paper. 
The consequence is that, within the approach presented in this work,  gauge-invariant observables (the Dirac observables) are not just simple combinations of the perturbation coordinates on the phase space even up to the first order. 
Having said this, in the following we may use the terminology ''gauge-invariant variables'' for those coordinates that commute with $E$, $M$, $V$, and $W$.  One should however bear in mind that they \emph{are not} gauge-invariant in the \emph{full} theory, and in fact we will be carrying out gauge-fixing in order for them to represent observable quantities.
\section{The constraints up to the first order: solutions and gauge-fixing} \label{solofconstruno}
\subsection{Solution to the Constraints} 
In this subsection, we solve the constraints up to the first order. Every solution to the full constraint coincides with one of our solutions up to the second (or higher) order in the perturbation variables.  Specifically, we will show the general solution to all the constraints, thus reducing to the constraint surface $\Gamma_C \subset \Gamma$. Next, we will choose a family of slices of $\Gamma_C$ transversal to the orbits of the gauge group generated by the constraints. We will finally find field variables freely parametrizing the slices and use them to fix the gauge, thus reducing to the physical phase space $\Gamma_{\rm phys}$.

To strat with, the constraint (\ref{C(k=0)}) up to the first order reads
\begin{equation} \label{uselss1}
\dfrac{1}{2} (p_T^{(0)})^2 + e^{6\alpha} V_T(T^{(0)}) - \dfrac{\kappa}{12} \pi_{\alpha}^2\ =\ 0
\end{equation}
We solve it with respect to $p_T^{(0)}$, finding
\begin{equation} \label{uselss2}
p_T^{(0)}\ =\ \pm \sqrt{\dfrac{\kappa}{6} \pi_\alpha^2 - 2 e^{6\alpha} V_T(T^{(0)})}
\end{equation}
The gauge transformations generated by $\breve C(0)$ can be used to fix arbitrarily the value of $T^{(0)}$:
\begin{equation} \label{uselss3}
T^{(0)}\ - \tau\ =\ 0
\end{equation} 
\\
The vector constraint equations
\begin{align} \label{vectorCeqs}
\left\{
\begin{array}{ll}
V(k) & = 0
\\
\\
W(k) & = 0
\end{array}
\right. \ \ \ \ \ \mbox{for all} \ k \in \mathcal{L} - \{0\}
\end{align}
can be immediately solved for the momenta $p^3$ and $p^4$:
\begin{align} \label{vectorCsol}
p^3(k) = -\dfrac{\pi_\alpha e^{-4\alpha}}{6} q_3(k), \ \ \ \ \ p^4(k) = -\dfrac{\pi_\alpha e^{-4\alpha}}{6} q_4(k)
\end{align}
The associated configuration variables, $(q_3(k), q_4(k))$, are free. They can be used to parametrize the gauge-orbits of the constraints $v^a C_a(k)$ and $w^a C_a(k)$ (which, at linear order, are nothing but $V(k)$ and $W(k)$) in the constraint surface $\Gamma_C \subset \Gamma$. 
\\
Let us now consider the scalar constraint equations:
\begin{align} \label{scalarCeqs}
\left\{
\begin{array}{ll}
E(k) & = 0
\\
\\
M(k) & = 0
\end{array}
\right. \ \ \ \ \ \mbox{for all} \ k \in \mathcal{L} - \{0\}
\end{align}
We proceed as above: we simply solve them for the momenta $p^1$ and $p^2$. The result is the following:
\begin{align} \label{scalarCsol}
\begin{array}{rcl}
p^1(k) & = & \dfrac{3 e^{4\alpha}}{\kappa \pi_\alpha} V_T' \delta \breve{T}(k) + \dfrac{3 p_T^{(0)} e^{-2\alpha}}{\kappa \pi_\alpha} \delta \breve{p}_T(k) -
\\
\\
& & - \left[\dfrac{\pi_\alpha e^{-4\alpha}}{8} + \dfrac{3 k^2}{\kappa^2 \pi_\alpha} + \dfrac{9 (p_T^{(0)})^2 e^{-4\alpha}}{4 \kappa \pi_\alpha} - \dfrac{9 e^{2\alpha}}{2 \kappa \pi_\alpha} V_T\right] q_1(k) + \dfrac{k^2}{\kappa^2 \pi_\alpha} q_2(k)
\\
\\
p^2(k) & = & \left[\dfrac{p_T^{(0)} e^{-2\alpha}}{2} - \dfrac{e^{4\alpha}}{\kappa \pi_\alpha} V_T'\right] \delta \breve{T}(k) - \dfrac{p_T^{(0)} e^{-2\alpha}}{\kappa \pi_\alpha} \delta \breve{p}_T(k) +
\\
\\
& & + \left[\dfrac{\pi_\alpha e^{-4\alpha}}{8} + \dfrac{k^2}{\kappa^2 \pi_\alpha} - \dfrac{3 (p_T^{(0)})^2 e^{-4\alpha}}{4 \kappa \pi_\alpha} - \dfrac{3 e^{2\alpha}}{2 \kappa \pi_\alpha} V_T\right] q_1(k) - \left[\dfrac{k^2}{3 \kappa^2 \pi_\alpha} + \dfrac{\pi_\alpha e^{-4\alpha}}{9}\right] q_2(k)
\end{array}
\end{align}
In this way,  $(q_1(k), q_2(k), \delta \breve{T}(k), \delta \breve{p}_T(k))$ can be used as free coordinates on the constraint surface $\Gamma_C$. Two of these free variables should be chosen to parametrize the gauge-orbits of $\breve C(k)$ and $k^a\breve C_a(k)$. The remaining two functions will represent the ''physical'' degrees of freedom.
\\
Finally, as for the tensor sector, we notice that it is completely unconstrained. Therefore, all variables $(q_5(k), p^5(k), q_6(k), p^6(k))$ are free: they are the ''physical'' degres of freedom associated with the graviton.
\\
The remaining variables (namely, the traceless matrices $\delta \breve q_{ab}(0)$ and $\delta \breve \pi^{ab}(0)$, as well as all the modes $\delta \phi(k)$ and $\delta\pi_\phi(k)$) are also free.

Of course all the ``free'' functions listed above are still subject to the gauge transformations, hence they represent the physical degrees of freedom in a gauge-dependent way. 
		\subsection{Gauge-Fixing}
At this point, we are following the so-called \emph{Reduced Phase Space} formalism, in which one first solves the constraints, and then identifies each gauge-orbit with a point in the physical phase space. This procedure is generally regarded as ideal, but can be implemented explicitly only in few cases. Indeed, it is usually a hard task to identify a set of gauge-invariant functions to coordinatize the physical phase space, and moreover they have usually a very complicated Poisson algebra, which makes canonical quantization practically impossible.
\\  
A way out of this problem is the so-called \emph{gauge-fixing} procedure. One chooses a set of functions to play the role of gauge-parameters, i.e. parametrizing the gauge-orbits of the constraints. Points along the same orbit are physically equivalent, so we can simply choose one to represent that specific orbit (physical state of the system). This amounts to fix the value of the gauge-parameters, choosing a ''slice'' in the constraint surface which mirrors the physical phase space. This slice is endowed with a symplectic form by simply pulling back the kinematical symplectic form along the embedding, and hence it is a good phase space. It is coordinatized by the remaining variables, which thus represent the physical degrees of freedom of the system. More precisely, the interpretation of these surviving variables is the following. We imagine to be dealing with Dirac-observables (i.e., functions that commute with all the constraints), denoted $\mathcal{O}_{\gamma, G}$. Now, $\gamma^I$ is the value of $\mathcal{O}_{\gamma, G}$ when restricting to the chosen slice (here collectively denoted by the gauge-fixing conditions $G_\alpha = 0$). There is a well developed formalism \cite{1, 2, 3, 4, 5, 6, 7, 8, 9} to treat these objects, and we will be using part of it when studying the dynamics. For now, it is enough to know that gauge-fixing is a ''legal'' procedure to reduce to the physical phase space.

In our case, we have to deal with 4 constraints $\breve C(k), \breve C_a(k)$ for each $k \neq 0$ and with $\breve C(0)$. Good gauge-parameters for such constraints can be chosen to be $(q_1(k), q_2(k), q_3(k), q_4(k), T^{(0)})$. The gauge-fixing conditions that we choose are the following:
\begin{eqnarray}
& \label{gaugeFix1} q_1(k) = q_2(k) = q_3(k) = q_4(k) = 0, \ \ \ \ \ \mbox{for all} \ k \in \mathcal{L} - \{0\} &
\\ \notag
\\
& \label{gaugeFix2} T^{(0)} - \tau = 0, \ \ \ \ \ \mbox{with} \ \tau \in \mathbb{R} &
\end{eqnarray}
Note that the value $\tau$ of gauge-parameter $T^{(0)}$ is left free: it will be used to label the Dirac observables and its changes will be used to describe their dynamics.
\section{The physical phase space, observables and their dynamics} \label{dinamo}
\subsection{Physical phase space and observables}
Mathematically,  the physical phase space $\Gamma_{\rm phys}$ for the theory we are considering in this paper is 
the space of the orbits in the constraint surface $\Gamma_C$ of the gauge transformation group generated by the scalar constraints  $C(x)$ and  the vector constraints $C_a(x)$.
The space $\Gamma_{\rm phys}$ can be embedded in $\Gamma_C$ as a slice which intersects each orbit exactly once. 
This is exactly what we did up to the first order in the previous section 
by solving the constraints and fixing a gauge.  Denote the image of the embedding by
$\Gamma_{\rm phys}^\tau$, that is 
\begin{align} \label{embedding}
\Gamma_{\rm phys} \rightarrow \Gamma_{\rm phys}^\tau \subset \Gamma_C\subset\Gamma
\end{align}
where $\tau$ is the  parameter used in the gauge  conditions.
The surface $\Gamma_{\rm phys}^\tau$ can be parametrised by the following variables originally defined in all the kinematical phase space $\Gamma$, 
\begin{align} \label{PSphysicalDOFs}
(\gamma_I) = \left(\alpha, \pi_\alpha, \delta \breve{q}_{ab}(0), \delta \breve{\pi}^{ab}(0), \delta \breve{\phi}(0), \delta \breve{\pi}_\phi(0), \delta \breve{T}(k), \delta \breve{p}_T(k), q_5(k), p^5(k), q_6(k), p^6(k), \delta \breve{\phi}(k), \delta \breve{\pi}_\phi(k)\right), \\
\mbox{for all} \ k \in \mathcal{L} - \{0\} \notag
\end{align}
The  embedding  (\ref{embedding})  determines  the surface $\Gamma_{\rm phys}^\tau$  in $\Gamma$ up to the 1st order as follows:
\begin{align} \label{PSnonphysicalDOFs}
\begin{array}{c}
q_1(k) = q_2(k) = q_3(k) = q_4(k) = p^3(k) = p^4(k) = 0
\\
\\
p^1(k) = \dfrac{3 e^{-2\alpha}}{\kappa \pi_\alpha} \left(p_T^{(0)} \delta \breve{p}_T(k) + e^{6\alpha} V_T' \delta \breve{T}(k)\right), \ \ \ \ \ p^2(k) = - \dfrac{p_T^{(0)} e^{-2\alpha}}{\kappa \pi_\alpha} \delta \breve{p}_T(k) + e^{-2\alpha} \left(\dfrac{p_T^{(0)}}{2} - \dfrac{e^{6\alpha}}{\kappa \pi_\alpha} V_T'\right) \delta \breve{T}(k)
\\
\\
T^{(0)} = \tau, \ \ \ \ \ p_T^{(0)} = \pm \sqrt{\dfrac{\kappa}{6} \pi_\alpha^2 - 2 e^{6\alpha} V_T(T^{(0)})}
\end{array}
\end{align}

The pullback of the coordinates (\ref{PSphysicalDOFs}) to $\Gamma_{\rm phys}$ 
defines coordinates 
\begin{align} \label{PSphysicalDOFstau}
(\gamma_I^\tau) = \left(\alpha^\tau, \pi_\alpha^\tau, \delta \breve{q}_{ab}(0)^\tau, \delta \breve{\pi}^{ab}(0)^\tau, \delta \breve{\phi}(0)^\tau, \delta \breve{\pi}_\phi(0)^\tau, \delta \breve{T}(k)^\tau, \delta \breve{p}_T(k)^\tau, q_5(k)^\tau, p^5(k)^\tau, q_6(k)^\tau, p^6(k)^\tau, \delta \breve{\phi}(k)^\tau, \delta \breve{\pi}_\phi(k)^\tau\right), \\
\mbox{for all} \ k \in \mathcal{L} - \{0\} \notag
\end{align}
on $\Gamma_{\rm phys}$. 
Note that, as emphasised by our notation, each of the functions $\gamma_I^\tau$ depends on the fixed value of $\tau$. We will come back to that dependence below, while defining the dynamics.  

What is independent of the embedding (\ref{embedding}) is the symplectic form $\Omega_{\rm phys}$, i.e., the pullback  to $\Gamma_{\rm phys}$  of the symplectic form in $\Gamma$.  This is the physical symplectic form.  To explicitely find it, we simply pull back the kinematical symplectic form
\begin{align} \label{Ksympl}
\Omega & = d\alpha \wedge d\pi_\alpha + dT^{(0)} \wedge dp_T^{(0)} + \dfrac{1}{2} d\delta \breve{q}_{ab}(0) \wedge d\delta \breve{\pi}^{ab}(0) + d\delta \breve{\phi}(0) \wedge d\delta \breve{\pi}_\phi(0) + \notag
\\
& + \sum_{k \in \mathcal{L} - \{0\}} \left[d\delta \breve{T}(k) \wedge d\delta \breve{p}_T(k) + \sum_{m = 1}^6 dq_m(k) \wedge dp^m(k) + d\delta \breve{\phi}(k) \wedge d\delta \breve{\pi}_\phi(k)\right]
\end{align}
Owing to  the convinient choice of gauge-fixing conditions, the pullbacks 
of $dT^{(0)}$, $dq_1$, ..., $dq_4$ by (\ref{embedding}) vanish identically. Therefore the pullback of the symplectic form reads simply
\begin{align} \label{Rsympl}
\Omega_{\rm phys} & = d\alpha^\tau \wedge d\pi_\alpha^\tau + \dfrac{1}{2} d\delta \breve{q}_{ab}(0)^\tau \wedge d\delta \breve{\pi}^{ab}(0){}^\tau + d\delta \breve{\phi}(0)^\tau \wedge d\delta \breve{\pi}_\phi(0)^\tau + \notag
\\
& + \sum_{k \in \mathcal{L} - \{0\}} \left[d\delta \breve{T}(k)^\tau \wedge d\delta \breve{p}_T(k)^\tau + \sum_{m = 5}^6 dq_m(k)^\tau \wedge dp^m(k)^\tau + d\delta \breve{\phi}(k)^\tau \wedge d\delta \breve{\pi}_\phi(k)^\tau\right]
\end{align}
The Poisson algebra that $\Omega_{\rm phys}$ defines on $\Gamma_{\rm phys}$ is easily found:
\begin{align} \label{PSphysicalALGEBRA}
\begin{array}{c}
\{\alpha^\tau, \pi_\alpha^\tau\}_{\rm phys} = 1, \ \ \ \ \ \{\delta \breve{q}_{ab}(0)^\tau, \delta \breve{\pi}^{cd}(0){}^\tau\}_{\rm phys} = \delta^c_{(a} \delta^d_{b)}- \frac{1}{3}\delta^{cd}\delta_{ab}, \ \ \ \ \ \{\delta \breve{\phi}(0)^\tau, \delta \breve{\pi}_\phi(0)^\tau\}_{\rm phys} = 1
\\
\\
\{q_5(k)^\tau, p^5(k')^\tau\}_{\rm phys} = \delta_{k, k'}, \ \ \ \ \ \{q_6(k)^\tau, p^6(k'){}^\tau\}_{\rm phys} = \delta_{k, k'}
\\
\\
\{\delta \breve{T}(k)^\tau, \delta \breve{p}_T(k')^\tau\}_{\rm phys} = \delta_{k, k'}, \ \ \ \ \ \{\delta \breve{\phi}(k)^\tau, \delta \breve{\pi}_\phi(k')^\tau\}_{\rm phys} = \delta_{k, k'}
\end{array}
\end{align}
This is the canonical Poisson algebra, the simplest we could hope to obtain: encouraging fact, in light of the future canonical quantization. 

Thus, we conclude that the reduced (physical) phase space $\Gamma_{\rm phys}$ is coordinatized by the functions (\ref{PSphysicalDOFstau}). 
Each of the variables (\ref{PSphysicalDOFstau}) defines a Dirac observable. Conversely, every {}{Dirac observable can be represented by a function $f(\gamma^\tau_I)$}.
The physical Poisson bracket between two such observales can be calculated from (\ref{PSphysicalALGEBRA}). This concludes the characterization of  the kinematical structure of the physical degrees of freedom of the theory. 
\subsection{Dynamics in $\Gamma_{\rm phys}$}
The dynamics of the theory is encoded in the dependence of the variables (\ref{PSphysicalDOFstau}) parametrising ${\Gamma_{\rm phys}}$ on the gauge parameter $\tau$.  Since the Poisson algebra 
(\ref{PSphysicalALGEBRA}) is canonical for every $\tau$, the dependence of the variables on $\tau$ is a flow of canonical transformations generated by some $\tau$-dependent function $h_{\rm phys}^\tau$ defined on $\Gamma_{\rm phys}$, and such that
\begin{equation}
\frac{d}{d\tau}\gamma_I^\tau\ =\ \{\gamma_I^\tau,h_{\rm phys}^\tau\}_{\rm phys}
\end{equation}
We call this function a \emph{physical Hamiltonian} \cite{fisarm1, fisarm2}. Obviously, it does not have the form of the canonical Hamiltonian $\int d^3x N(x)C(x)+N^a(x)C_a(x)$, because the canonical Hamiltonian vanishes identically on $\Gamma_C$ in which $\Gamma_{\rm phys}$ is embedded. On the other hand, $h_{\rm phys}$ must follow somehow from the canonical dynamics $\int d^3x N(x)C(x)+N^a(x)C_a(x)$. Therefore, to derive the physical Hamiltonian we go back to the constraint surface $\Gamma_C$ in the kinematical phase space $\Gamma$, and even to $\Gamma$ itself (because $\Gamma_C$ is not equipped with a symplectic form or with the Poisson bracket).

Let us use the projection 
\begin{align} \label{CprojectionC}
\Pi:\Gamma_C\rightarrow \Gamma_{\rm phys}
\end{align}
to pullback every function $f$ defined on $\Gamma_{\rm phys}$, to a function 
on $\Gamma_C$, 
\begin{equation} \label{Pi*}
 {\cal O}_f\ =\ \Pi^*f
 \end{equation}
called Dirac observable. It is constant on each orbit of the gauge transformations, and extended arbitrarily to the kinematical phase space $\Gamma$. This Dirac observable 
${\cal O}_f$ weakly Poisson-commutes with the constraints, so in particular
\begin{equation} \label{{O,C}}
\left.\left\{{\cal O}_f, \int d^3x N(x)C(x)+N^a(x)C_a(x)\right\}\right|_{\Gamma_C}\ =\ 0
\end{equation} 
Conversely, every Dirac observable defines a function $f$ on $\Gamma_{\rm phys}$. 
The kinematical Poisson algebra of the Dirac observables corresponding to the variables parametrising the physical phase space $\Gamma_{\rm phys}$ is consistent with the physical Poisson algebra in $\Gamma_{\rm phys}$,  that is
\begin{equation}\label{{O,O}}
\{{\cal O}_f,{\cal O}_{f'}\}|_{\Gamma_C}\ =\ {\cal O}_{\{f,f'\}_{\rm phys}}\,{}|_{\Gamma_C}
\end{equation} 

At this point, let us choose a trivial potential $V_T$ for the theory (\ref{action}):
\begin{equation} \label{trivvi}
V_T = 0
\end{equation}
It follows that the theory is invariant with respect to the translation
\begin{equation} \label{TtranslationT}
T\mapsto T+\tau
\end{equation}
Therefore, if ${\cal O}$ is a Dirac observable, so is the following function ${\cal O}^\tau$:
\begin{equation} \label{TtranslatedO}
{\cal O}^\tau(T,...)\ :=\ {\cal O}(T-\tau,...)
\end{equation}
where ``...'' stands for the remaining variables which are the same on the both sides of the equality. In particular, given $\gamma_I^{\tau}$, and $\gamma_I^{\tau+\Delta\tau}$,
the corresponding Dirac observables ${\cal O}_{\gamma_I^\tau}$ and ${\cal O}_{\gamma_I^{\tau+\Delta\tau}}$ are related as follows,
\begin{equation} \label{almostThere1}
{\cal O}_{\gamma_I^{\tau+\Delta\tau}}(T,...)\ =\ {\cal O}_{\gamma_I^\tau}(T-\Delta\tau,...)
\end{equation}
It follows that
\begin{equation} \label{almostThere2}
\frac{d}{d\tau}{\cal O}_{\gamma_I^{\tau}} \ =\ -\frac{\partial}{\partial T^{(0)}}{\cal O}_{\gamma_I^\tau}
\end{equation}
But the right hand side can be calculated from equation (\ref{{O,C}}), where we have still freedom in  choosing the lapse function and the shift vector.
{}{To find the most convenient choice, let us write
\begin{equation} \label{decomp}
C(x) = \dfrac{1}{2\sqrt{q(x)}} (p_T(x)^2 - h(x)^2) = \dfrac{p_T(x) + h(x)}{2\sqrt{q(x)}} (p_T(x) - h(x))
\end{equation}
(we can always do so, since $C(x) = 0$ and $p_T^2/2\sqrt{q} \geq 0$ imply that $C(x) - p_T^2/2\sqrt{q} \leq 0$, and can then be written as the negative of a square). The function $h(x)^2$ can be explicitely calculated from the scalar constraint, and it has the form
\begin{align} \label{dyn-hsq}
h^2 = - 4\kappa \left(\pi_{ab} \pi^{ab} - \dfrac{1}{2}(q_{ab} \pi^{ab})^2\right) + \dfrac{1}{\kappa} q R^{(3)} - \pi_\phi^2 - q q^{ab} \partial_a \phi \partial_b \phi - 2q V_\phi(\phi) - q q^{ab} \partial_a T \partial_b T
\end{align}
It is immediate to see that $h^2$ (and hence its square root $h$) does not involve $T^{(0)}$ nor $p_T^{(0)}$:
\begin{align} \label{clockIndependence}
\frac{\partial }{\partial T^{(0)}}h(x)\ =\ 0, \ \ \ \ \ \frac{\partial }{\partial p_T^{(0)}}h(x)\ =\ 0
\end{align}
Now, a good choice for lapse and shift is\footnote{
Another choice for the lapse function would be $N = 2\sqrt{q}/(p_T - h)$. In this case, $H = p_T^{(0)} + \tilde{h}$. But since $\tilde{h} \geq 0$, from $H = 0$ it would follow that $p_T^{(0)} \leq 0$. This is mathematically acceptable (it corresponds to a contracting FRW universe, but is not the physical universe in which we live (which is expanding, hence $p_T^{(0)} \geq 0$).
}
\begin{align} \label{aGoodC}
N = 2\sqrt{q}/(p_T + h), \ \ \ \ \ N^a = 0
\end{align}
In this way, the canonical Hamiltonian $H$ present in equation (\ref{{O,C}}), reduces to
\begin{align} \label{dyn-true}
H = \int d^3x (p_T(x) - h(x)) = p_T^{(0)} - \tilde{h}
\end{align}
having used the fact that $\int d^3x p_T(x) = p_T^{(0)}$ and having defined 
\begin{align} \label{dyn-Hphys}
\tilde{h} := \int d^3x h(x) \ =\ \Pi^*h_{\rm phys}
\end{align}
}
From all this, it follows that
\begin{equation}
0\ =\ \{{\cal O}_{\gamma_I^{\tau}},p_T^{(0)} - \tilde{h}\}\ =\  \frac{\partial}{\partial T^{(0)}}{\cal O}_{\gamma_I^{\tau}} - \{{\cal O}_{\gamma_I^{\tau}},\tilde{h}\}
\end{equation}
Note that $p_T^{(0)}$ is a Dirac observable. But (\ref{dyn-true}) implies that so is $\tilde{h}$, and hence 
\begin{equation}
\tilde{h}\ =\ {\cal O}_{h_{\rm phys}}
\end{equation}   
where $h_{\rm phys}$ is a function defined on $\Gamma_{\rm phys}$ by $\tilde{h}$.
Finally, we have
\begin{equation}
\frac{d}{d\tau}{\cal O}_{\gamma_I^{\tau}}\ =\ -O_{\{\gamma_I^{\tau},h_{\rm phys}\}_{\rm phys}} 
\end{equation}
Purpose of the next subsection is to explicitely express the physical Hamiltonian as a function of the free coordinates $(\gamma^\tau_I)$ on the physical phase space:
\begin{equation} \label{hphys2}
h_{\rm phys}(\gamma^\tau_I)\ =\ \tilde{h}\left(\gamma_I^\tau, q_n = 0, T^{(0)} = \tau, p^n = p^n(\gamma_I^\tau), p_T^{(0)} = p_T^{(0)}(\gamma_I^\tau)\right)
\end{equation} 
where $p^n(\gamma_I^\tau),...$ (with $n = 1, 2, 3, 4$) is given by dropping the subscripts $\tau$, {}{using  (\ref{PSnonphysicalDOFs}) and restoring the subscripts $\tau$ again.}
\subsection{Explicit form of the physical Hamiltonian}
The derivation in the previous section -- a self-contained construction which is a special case of the powerfull theory of relational observables \cite{1, 2, 3, 4, 5, 6, 7} -- is exact. However, to explicitely express (\ref{hphys2}), we need to go back to the expansion in the perturbation variables.

Let us consider the argument (\ref{dyn-hsq}) of the square root. Plugging the expansions (\ref{perts2}) in it, we find
\begin{align} \label{dyn-hargument}
h^2 & = - 4\kappa \left[(e^{2\alpha} \delta_{ac} + \delta q_{ac}) (e^{2\alpha} \delta_{bd} + \delta q_{bd}) \left(\dfrac{\pi_\alpha}{6} e^{-2\alpha} \delta^{ab} + \delta \pi^{ab}\right) \left(\dfrac{\pi_\alpha}{6} e^{-2\alpha} \delta^{cd} + \delta \pi^{cd}\right) \right. - \notag
\\
& - \left. \dfrac{1}{2} \left((e^{2\alpha} \delta_{ab} + \delta q_{ab}) \left(\dfrac{\pi_\alpha}{6} e^{-2\alpha} \delta^{ab} + \delta \pi^{ab}\right)\right)^2\right] + \dfrac{1}{\kappa} (\delta^{(0)}q + \delta^{(1)}q + \delta^{(2)}q) (\delta^{(0)}R^{(3)} + \delta^{(1)}R^{(3)} + \delta^{(2)}R^{(3)}) - \notag
\\
& - \delta \pi_\phi^2 - (\delta^{(0)}q + \delta^{(1)}q + \delta^{(2)}q) \left[(\delta^{(0)}q^{ab} + \delta^{(1)}q^{ab} + \delta^{(2)}q^{ab}) \left(\partial_a \phi \partial_b \phi + \partial_a T \partial_b T\right) + 2 V_\phi(\phi)\right]
\end{align}
Here, we denoted by $\delta^{(i)}q$, $\delta^{(i)}R^{(3)}$ and $\delta^{(i)}q^{ab}$ the $i$th order of $q$, $R^{(3)}$ and $q^{ab}$ respectively. The expansions of such phase space functions can be found in the appendix.

Expanding these products and keeping up to $2$nd order, one finally groups the various terms according to their order. Formally, $h^2$ is of the form
\begin{align} \label{dyn-Hfor}
h^2 = A + B\epsilon + C\epsilon'
\end{align}
where $\epsilon$ and $\epsilon'$ are respectively linear and quadratic in the perturbation variables. Explicitely,
\begin{align}
\label{dyn-Hfor1} A & = \dfrac{\kappa \pi_\alpha^2}{6}
\\ \notag
\\
\label{dyn-Hfor2} B \epsilon & = \dfrac{\kappa \pi_\alpha^2 e^{-2\alpha}}{9} \delta^{ab} \delta q_{ab} + \dfrac{2 \kappa \pi_\alpha e^{2\alpha}}{3} \delta_{ab} \delta \pi^{ab} + \dfrac{e^{2\alpha}}{\kappa} \partial^a \partial^b \delta q_{ab} - \dfrac{e^{2\alpha}}{\kappa} \delta^{ab} \partial_e \partial^e \delta q_{ab}
\\ \notag
\\
\label{dyn-Hfor3} C \epsilon' & = - \dfrac{\kappa \pi_\alpha^2 e^{-4\alpha}}{9} \left(\delta^{ab} \delta^{cd} - \dfrac{1}{2} \delta^{ac} \delta^{bd}\right) \delta q_{ac} \delta q_{bd} - 4\kappa e^{4\alpha} \left(\delta_{ab} \delta_{cd} - \dfrac{1}{2} \delta_{ac} \delta_{bd}\right) \delta \pi^{ac} \delta \pi^{bd} - \notag
\\
& - \dfrac{2 \kappa \pi_\alpha}{3} \delta q_{ab} \delta \pi^{ab} + \dfrac{2 \kappa \pi_\alpha}{3} \delta^{ab} \delta_{cd} \delta q_{ab} \delta \pi^{cd} + \dfrac{1}{\kappa} \delta^{ab} \delta q_{ab} \partial^c \partial^d \delta q_{cd} - \dfrac{1}{\kappa} \delta^{ab} \delta q_{ab} \delta^{cd} \partial_e \partial^e \delta q_{cd} - \notag
\\
& - \dfrac{2}{\kappa} \delta^{ab} \delta q_{bc} \partial^c \partial^d \delta q_{da} + \dfrac{1}{\kappa} \delta^{ac} \delta^{bd} \delta q_{ab} \partial_e \partial^e \delta q_{cd} + \dfrac{1}{\kappa} \delta q_{ab} \delta^{cd} \partial^a \partial^b \delta q_{cd} + \dfrac{1}{\kappa} \delta^{cd} \partial^a \delta q_{ab} \partial^b \delta q_{cd} - \notag
\\
& - \dfrac{1}{4\kappa} \delta^{ab} \delta^{cd} \partial_e \delta q_{ab} \partial^e \delta q_{cd} - \dfrac{1}{2\kappa} \delta^{ab} \partial^d \delta q_{ac} \partial^c \delta q_{bd} - \dfrac{1}{\kappa} \delta^{cd} \partial^a \delta q_{ac} \partial^b \delta q_{bd} + \dfrac{3}{4\kappa} \delta^{ac} \delta^{bd} \partial_e \delta q_{ab} \partial^e \delta q_{cd} - \notag
\\
& - \delta \pi_\phi^2 - e^{4\alpha} \delta^{ab} \partial_a \delta \phi \partial_b \delta \phi - e^{6\alpha} V_\phi''(0) \delta \phi^2 - e^{4\alpha} \delta^{ab} \partial_a \delta T \partial_b \delta T
\end{align}
Now, recall the Taylor expansion of a square root of two variables:
\begin{align} \label{dyn-taylor}
\sqrt{A + B\epsilon + C\epsilon'} = \sqrt{A} + \dfrac{B\epsilon}{2\sqrt{A}} + \dfrac{1}{2 \sqrt{A}} \left(C \epsilon' - \dfrac{(B\epsilon)^2}{4A}\right)
\end{align}
Thus, we can write $h_\text{phys}$ as
\begin{align} \label{dyn-HPexp}
h_\text{phys} = h_\text{phys}^{(0)} + h_\text{phys}^{(1)} + h_\text{phys}^{(2)} = \sqrt{A} + \dfrac{1}{2\sqrt{A}} \int d^3x B\epsilon + \dfrac{1}{2 \sqrt{A}} \int d^3x \left(C \epsilon' - \dfrac{(B\epsilon)^2}{4A}\right)
\end{align}
Let us consider the three orders separately.

The $0$th order, $h_\text{phys}^{(0)}$, corresponds to the homogeneous Hamiltonian
\begin{align} \label{dyn-HP0}
h_\text{phys}^{(0)} = \sqrt{\dfrac{\kappa (\pi_\alpha^\tau)^2}{6}} =: H_\text{hom}
\end{align}
i.e. the Hamiltonian that generates the dynamics for the geometry in the case no perturbations are considered (FRW spacetime).

As for the $1$st order, notice that it involves an integral over the whole space of objects which are linear in the perturbations: it is not a surprise that, once we Fourier-transform it, it vanishes identically.

The $2$nd order is thus the first correction to the dynamics. After some algebra, and using the simple rules presented in the appendix for dealing with the Fourier transform, we obtain an object of the form
\begin{align} \label{dyn-HP2}
h_\text{phys}^{(2)} = \sqrt{\dfrac{6}{\kappa (\pi_\alpha^\tau)^2}} \left[D(0) + \sum_{k \in \mathcal{L} - \{0\}}D(k)\right]
\end{align}
where $D(0)$ contains $\delta \breve{q}_{ab}(0)^\tau$ and $\delta \breve{\phi}(0)^\tau$ (and their conjugate momenta), while $D(k)$ contains the $k \neq 0$ modes, which can be expanded as $\delta \breve{q}_{ab}(k)^\tau = A(k)_{ab}^m q_m^\tau$ (and same for its momentum). Using the properties of all these objects, and imposing the constraints and the gauge conditions, one finds that on the physical phase space $\Gamma_{\rm phys}$ it is
\begin{align}
\label{dyn-HPD1} D(0) & = - 2 \kappa e^{4\alpha^\tau} \delta_{ab} \delta_{cd} \delta \breve{\pi}^{ac}(0)^\tau \delta \breve{\pi}^{bd}(0)^\tau - \dfrac{\kappa (\pi_\alpha^\tau)^2 e^{-4\alpha^\tau}}{18} \delta^{ab} \delta^{cd} \delta \breve{q}_{ac}(0)^\tau \delta \breve{q}_{bd}(0)^\tau - \dfrac{\kappa \pi_\alpha^\tau}{3} \delta \breve{q}_{ab}(0)^\tau \delta \breve{\pi}^{ab}(0)^\tau - \notag
\\
& - \dfrac{1}{2} \left[(\delta \breve{\pi}(0)^\tau)^2 + e^{6\alpha^\tau} V_\phi''(0) (\delta \breve{\phi}(0)^\tau)^2\right]
\\
\\
\label{dyn-HPD2} D(k) & = \sum_{m = 5, 6} \left[- 2 \kappa e^{4\alpha^\tau} (p^m(k)^\tau)^2 - \dfrac{\kappa \pi_\alpha^\tau}{3} q_m(k)^\tau p^m(k)^\tau - \dfrac{\kappa (\pi_\alpha^\tau)^2 e^{-4\alpha^\tau}}{18} (q_m(k)^\tau)^2 - \dfrac{k^2}{8 \kappa} (q_m(k)^\tau)^2\right] - \notag
\\
& - \dfrac{1}{2} \left[(\delta \breve{p}_T(k)^\tau)^2 - \kappa \pi_\alpha^\tau \delta \breve{T}(k)^\tau \delta \breve{p}_T(k)^\tau + \dfrac{\kappa^2 (\pi_\alpha^\tau)^2}{4} (\delta \breve{T}(k)^\tau)^2 + e^{4\alpha^\tau} k^2 (\delta \breve{T}(k)^\tau)^2\right] - \notag
\\
& - \dfrac{1}{2} \left[(\delta \breve{\pi}(k)^\tau)^2 + e^{4\alpha^\tau} k^2 (\delta \breve{\phi}(k)^\tau)^2 + e^{6\alpha^\tau} V_\phi''(0) (\delta \breve{\phi}(k)^\tau)^2\right]
\end{align}

In this way, we have derived the explicit form of $h_\text{phys}$ (see (\ref{hphys2})) up to the $2$nd order. It is convenient to group the terms, according to their dependence on the fields:
\begin{align} \label{dyn-HP}
h_\text{phys} = H_\text{hom} + H_{k = 0} + \sum_{k \neq 0, m = 5, 6} H^G_{m, k} + \sum_{k \neq 0} H^T_k + \sum_k H^M_k
\end{align}
where
\begin{align} \label{dyn-tutte}
\begin{array}{lcl}
H_\text{hom} & = & \sqrt{\dfrac{\kappa (\pi_\alpha^\tau)^2}{6}}
\\
\\
H_{k = 0} & = & - \sqrt{\dfrac{6}{\kappa (\pi_\alpha^\tau)^2}} \left[2 \kappa e^{4\alpha^\tau} \delta_{ab} \delta_{cd} \left(\delta \breve{\pi}^{ac}(0)^\tau + \dfrac{\pi_\alpha^\tau e^{-4\alpha^\tau}}{12} \delta^{ae} \delta^{cf} \delta \breve{q}_{ef}(0)^\tau\right) \left(\delta \breve{\pi}^{bd}(0)^\tau + \dfrac{\pi_\alpha^\tau e^{-4\alpha^\tau}}{12} \delta^{bg} \delta^{dh} \delta \breve{q}_{gh}(0)^\tau\right) +\right.
\\
\\
& & \left.+ \dfrac{\kappa (\pi_\alpha^\tau)^2 e^{-4\alpha^\tau}}{24} \delta^{ab} \delta^{cd} \delta \breve{q}_{ac}(0)^\tau \delta \breve{q}_{bd}(0)^\tau\right]
\\
\\
H^G_{m, k} & = & - \sqrt{\dfrac{6}{\kappa (\pi_\alpha^\tau)^2}} \left[2 \kappa e^{4\alpha^\tau} \left(p^m(k)^\tau + \dfrac{\pi_\alpha^\tau e^{-4\alpha^\tau}}{12} q_m(k)^\tau\right)^2 + \dfrac{1}{2} \left(\dfrac{\kappa (\pi_\alpha^\tau)^2 e^{-4\alpha^\tau}}{12} + \dfrac{k^2}{4 \kappa}\right) (q_m(k)^\tau)^2\right]
\\
\\
H^T_k & = & - \sqrt{\dfrac{6}{\kappa (\pi_\alpha^\tau)^2}} \left[\dfrac{1}{2} \left(\delta \breve{p}_T(k)^\tau - \dfrac{\kappa \pi_\alpha}{2}^\tau \delta \breve{T}(k)^\tau\right)^2 + \dfrac{1}{2} e^{4\alpha^\tau} k^2 (\delta \breve{T}(k)^\tau)^2\right]
\\
\\
H^M_k & = & - \sqrt{\dfrac{6}{\kappa (\pi_\alpha^\tau)^2}} \left[\dfrac{(\delta \breve{\pi}(k)^\tau)^2}{2} + \dfrac{1}{2} \left(e^{4\alpha^\tau} k^2 + e^{6\alpha^\tau} V_\phi''(0)\right) (\delta \breve{\phi}(k)^\tau)^2\right]
\end{array}
\end{align}
can be thought of as the various Hamiltonians generating the dynamics on the different sectors of the physical phase space. This concludes our exposition of the physical dynamics of the theory.
\section{A remark about the Mukhanov-Sasaki variables} \label{MSwrongs}
This  is a good point to bridge our approach with the one used in the standard cosmological pertrubation theory (and, at the same time, showing why our is better suited in the context of quantum field theory on quantum cosmological spacetime). This fact is that, if we consider only the constraints $E(k),M(k),W(k),V(k)$ (that is the linear parts $\breve C(k)^{(1)},\breve C_a^{(1)}(k)$ of the constraints of the full theory), then  the gauge-fixing procedure we just presented is not necessary: indeed, the gauge-invariant scalar degrees of freedom are known, and are called \emph{Mukhanov-Sasaki variables} \cite{MandS}. In terms of our variables, they are
\begin{align} \label{g-invScalarModes}
\left\{
\begin{array}{ll}
Q(k) & = \delta \breve{T}(k) + \dfrac{3 p_T^{(0)} e^{-2\alpha}}{\kappa \pi_\alpha} q_1(k) - \dfrac{p_T^{(0)} e^{-2\alpha}}{\kappa \pi_\alpha} q_2(k)
\\
\\
P(k) & = \delta \breve{p}_T(k) - \dfrac{\kappa \pi_\alpha}{2} \delta \breve{T}(k) - 3 e^{-2\alpha} \left(\dfrac{p_T^{(0)}}{2} + \dfrac{e^{6\alpha}}{\kappa \pi_\alpha} V_T'(T^{(0)})\right) q_1(k) + \dfrac{e^{4\alpha}}{\kappa \pi_\alpha} V_T'(T^{(0)}) q_2(k)
\end{array}
\right.
\end{align}
They are used in standard cosmological perturbation theory, because $Q(k)$ and $P(k)$ commute with the linearized constraints and have the canonical Poisson algebra:
\begin{eqnarray}
\label{scalarGinvariance} & \{Q(k), E(k')\} = \{Q(k), M(k')\} = 0, \ \ \ \ \ \{P(k), E(k')\} = \{P(k), M(k')\} = 0 &
\\ \notag
\\
\label{scalarAlgebra} & \{Q(k), P(k')\} = \delta_{k, k'}, \ \ \ \ \ \{Q(k), Q(k')\} = \{P(k), P(k')\} = 0 &
\end{eqnarray}
Therefore,  one does not need to fix  a gauge, because the Dirac observables coordinatizing the physical phase space are already available: $(q_5(k), p^5(k), q_6(k), p^6(k), Q(k), P(k))$, with the canonical Poisson algebra.

However, there is a very important reason for us not to follow this route. Recall that in our approach the background variables  $\alpha,\pi_\alpha,T^{(0)}$ and $p_T^{(0)}$  are treated on the same footing as the perturbation variables (i.e., they are variables of the phase space, to be quantized). But (\ref{g-invScalarModes}) shows that $Q(k)$ and $P(k)$ are functions of the background variables, which means that whenever computing Poisson brackets involving them, one has to take into account the Poisson algebra of the background variables as well. Secondly, we have different constraints: namely, we also regard the contribution to gauge transformations coming from the second order terms of $\breve C(k)$ and $C_a(k)$, and we consider the constraint $\breve C(0)$. The variables $Q(k)$ and $P(k)$ \emph{are not} invariant with respect to all those gauge transformations.  For this reason, even though $Q(k)$ and $P(k)$ are perfectly good variables for studying quantized inhomogeneous perturbations on a curved \emph{fixed} classical background, they are not suited for the purpose of quantizing  the perturbations \emph{and}  the background simultanouesly. 

Nevertheless, it is interesting to note that, with respect to these variables, the dynamics seemingly simplifies a lot. Indeed, let uas go  back to (\ref{dyn-tutte}). There, we grouped the terms in such a way that the Hamiltonians take the  form of the Hamiltonian of the harmonic oscillator. More precisely: defining the new ''momenta''
\begin{align} \label{dyn-momenta}
\begin{array}{lll}
\delta \breve{\Pi}^{ab}(0)^\tau & := & \breve{\pi}^{ab}(0)^\tau + \dfrac{\pi_\alpha^\tau e^{-4\alpha^\tau}}{12} \delta^{ac} \delta^{bd} \delta \breve{q}_{cd}(0)^\tau
\\
\\
P^m(k)^\tau & := & p^m(k)^\tau + \dfrac{\pi_\alpha^\tau e^{-4\alpha^\tau}}{12} q_m(k)^\tau
\\
\\
\delta \breve{P}_T(k)^\tau & := & \delta \breve{p}_T(k)^\tau - \dfrac{\kappa \pi_\alpha^\tau}{2} \delta \breve{T}(k)^\tau
\end{array}
\end{align}
we see that (\ref{dyn-tutte}) reduce to
\begin{align} \label{dyn-tutteNEW}
\begin{array}{lcl}
H_\text{hom} & = & \sqrt{\dfrac{\kappa}{6}} \pi_\alpha^\tau
\\
\\
H_{k = 0} & = & - \sqrt{\dfrac{6}{\kappa (\pi_\alpha^\tau)^2}} \left[2 \kappa e^{4\alpha^\tau} \delta_{ab} \delta_{cd} \delta \breve{\Pi}^{ac}(0)^\tau \delta \breve{\Pi}^{bd}(0)^\tau + \dfrac{\kappa (\pi_\alpha^\tau)^2 e^{-4\alpha^\tau}}{24} \delta^{ab} \delta^{cd} \delta \breve{q}_{ac}(0)^\tau \delta \breve{q}_{bd}(0)^\tau\right]
\\
\\
H^G_{m, k} & = & - \sqrt{\dfrac{6}{\kappa (\pi_\alpha^\tau)^2}} \left[2 \kappa e^{4\alpha^\tau} (P^m(k)^\tau)^2 + \dfrac{1}{2} \left(\dfrac{\kappa (\pi_\alpha^\tau)^2 e^{-4\alpha^\tau}}{12} + \dfrac{k^2}{4 \kappa}\right) (q_m(k)^\tau)^2\right]
\\
\\
H^T_k & = & - \sqrt{\dfrac{6}{\kappa (\pi_\alpha^\tau)^2}} \left[\dfrac{(\delta \breve{P}_T(k)^\tau)^2}{2} + \dfrac{1}{2} e^{4\alpha^\tau} k^2 (\delta \breve{T}(k)^\tau)^2\right]
\\
\\
H^M_k & = & - \sqrt{\dfrac{6}{\kappa (\pi_\alpha^\tau)^2}} \left[\dfrac{(\delta \breve{\pi}(k)^\tau)^2}{2} + \dfrac{1}{2} \left(e^{4\alpha^\tau} k^2 + e^{6\alpha^\tau} V_\phi''(0)\right) (\delta \breve{\phi}(k)^\tau)^2\right]
\end{array}
\end{align}
From here, we see immediately how the various degrees of freedom behave:
\begin{itemize}
\item the (two polarizations of the) graviton behaves as a free relativistic particle with mass induced by the background geometry (via the term proportional to $(\pi_\alpha^\tau)^2$)
\item the perturbations of the clock field $T$ propagate as massless relativistic particles
\item the (perturbations of the) test field $\phi$ propagates as massive relativistic particle, where the mass is given by the second derivative of its potential $V_\phi$ (as it happens in flat spacetime)
\end{itemize}
In particular, $H_T$ is the Hamiltonian acting on the physical scalar sector, and using (\ref{g-invScalarModes}) on the reduced phase space, one sees that it reduces to
\begin{align} \label{dyn-MSh}
H^T_k = - \sqrt{\dfrac{6}{\kappa (\pi_\alpha^\tau)^2}} \left[\dfrac{(P(k)^\tau)^2}{2} + \dfrac{1}{2} e^{4\alpha^\tau} k^2 (Q(k)^\tau)^2\right]
\end{align}
In other words, if we stop at the linear order, our scalar sector (and its dynamics) coincides with the one found in standard cosmological perturbation theory.

However, it is important to observe that -- from the point of view of the full theory (which we are considering) -- the transformation (\ref{dyn-momenta}) is not canonical. In particular, as already pointed out, the Poisson algebra with the background geometry is nontrivial:
\begin{align} \label{dyn-nontrivial}
\begin{array}{llll}
\{\alpha^\tau, \delta \breve{\Pi}^{ab}(0)^\tau\} = \dfrac{e^{-4\alpha^\tau}}{12} \delta^{ac} \delta^{bd} \delta \breve{q}_{cd}(0)^\tau, & & & \{\pi_\alpha^\tau, \delta \breve{\Pi}^{ab}(0)^\tau\} = \dfrac{\pi_\alpha e^{-4\alpha^\tau}}{3} \delta^{ac} \delta^{bd} \delta \breve{q}_{cd}(0)^\tau
\\
\\
\{\alpha^\tau, P^m(k)^\tau\} = \dfrac{e^{-4\alpha^\tau}}{12} q_m(k)^\tau, & & & \{\pi_\alpha^\tau, P^m(k)^\tau\} = \dfrac{\pi_\alpha^\tau e^{-4\alpha^\tau}}{3} q_m(k)^\tau
\\
\\
\{\alpha^\tau, \delta \breve{P}_T(k)^\tau\} = -\dfrac{\kappa}{2} \delta \breve{T}(k)^\tau, & & & \{\pi_\alpha^\tau, \delta \breve{P}_T(k)^\tau\} = 0
\end{array}
\end{align}
We thus have a dilemma:  the simple form of the Hamiltonian is traded for a more complicated Poisson algebra, which mixes the pertubations to the clock field with the background geometry. It is important to realize this fact when carrying out the canonical quantization of linearized inhomogeneous modes \emph{and} of the homogeneous isotropic background! We think that a simple kinematics is a better starting point, and thus would choose the original momenta, rather than the new ones (\ref{dyn-momenta}).

Finally, notice that the test field variables $\delta \breve{\phi}^\tau$ and $\delta \breve{\pi}_\phi^\tau$ are real canonical variables (i.e., the only non-trivial Poisson brakets are $\{\delta \breve{\phi}(k)^\tau, \delta \breve{\pi}_\phi(k')^\tau\} = \delta_{k, k'}$). Therefore,  for each mode of the test field $\phi$ the Hamiltonian does have the canonical form of the harmonic oscillator. Due to this fact, the very result of our systematic analysis obtained in the current paper coincides with that of \cite{iso} obtained by using short cuts, after the restriction to $\phi$. 
\section{Summary, conclusions and outlook} \label{finel}
In this paper, we provided a  framework for quantization of linear perturbations (inhomogeneities) on a quantum background spacetime. We hope to have convinced the reader that, in light of canonical quantization, the classical Poisson algebra is the most fundamental feature to be preserved at the quantum level. This imposes a different choice of fundamental variables than that usually taken when quantizing perturbations on a fixed classical background spacetime. In particular, we showed that Mukhanov-Sasaki variables are not suited for this purpose. However, a natural gauge-fixing exists, which allows to use the old variables as fundamental operators, and provides a true dynamics in terms of the homogeneous part of the clock scalar field, $T$. Of course, it is expected that this formalism can be developed for other choices of physical time as well \cite{GandT1, GandT2, dust}.

{}{Technically, the goal of this work was derivation of the formulae (\ref{dyn-HP})-(\ref{dyn-tutte}) for the physical Hamiltonian $h_{\rm phys}$. Therein, the Hamiltonian is expressed  by the Dirac observables $\gamma_I^\tau$ (\ref{PSphysicalDOFstau}). Given value of the parameter $\tau$, the observables  parametrise  the phase space of solutions to the constraints modulo the gauge transformations.  The first two Dirac observables, $\alpha^\tau$ and $\pi_\alpha^\tau$, are identified with the background degrees of freedom.
The remaining Dirac observables are perturbations: $\delta \breve{q}_{ab}(0)^\tau, \delta \breve{\pi}^{ab}(0)^\tau, \delta \breve{\phi}(0)^\tau, \delta \breve{\pi}_\phi(0)^\tau, \delta \breve{T}(k)^\tau, \delta \breve{p}_T(k)^\tau, q_5(k)^\tau, p^5(k)^\tau, q_6(k)^\tau, p^6(k)^\tau, \delta \breve{\phi}(k)^\tau, \delta \breve{\pi}_\phi(k)^\tau$ where $ k \in \mathcal{L} - \{0\} \notag$.
Their Poisson algebra is canonical (\ref{PSphysicalALGEBRA}). It is defined  by the proper kinematical Poisson algebra of the full theory of the gravitational field coupled to two K-G fields.
The constraints are solved up to the first order, and the Hamiltonian itself is derived up to the second order in the perturbation variables. The physical Hamiltonian generates the dynamics via $d\gamma_I^\tau/d\tau = -\{\gamma_I^\tau,h_{\rm phys}\}$, derived up to the first order in the perturbation variables. The dynamics of the background degrees of freedom $\alpha^\tau,\pi_\alpha^\tau$ is an undecoupled  part of the dynamics of all the system parametrised by both the background variables and the variables we perturb with respect to. Going to higher orders in the perturbations amounts to simply adding to $h_{\rm phys}$ higher order terms in the perturbation variables and to imposing  so-called linearization-stability constraints linear in the perturbations \cite{LSC1, LSC2, LSC3, LSC4, LSC5, LSC6, LSC7, LSC8}.}
                
{}{The effective difference between our results and the results of the standard approach to cosmological perturbations consists in the status of the Mukhanov-Sasaki variables $Q(k)$ and $P(k)$: to begin with, they are not Dirac observables themselves in our approach, however, as any other function on the phase space they can be assigned Dirac observables $Q(k)^\tau$ and $P(k)^\tau$ in a $\tau$-dependent manner.  They still provide the corresponding term of the physical Hamiltonian with the canonical form (\ref{dyn-MSh}). Nonetheless, in our framework they do not Poisson-commute with the background degrees of freedom (see (\ref{dyn-nontrivial})). This last fact has to be taken into account in the process of quantization. The consequence is that, whereas according to the standard approach the perturbations of the K-G field $T$ non-vanishing in the background define the Hamiltonian term $H_T$ of the same form as the Hamiltonian term $H_M$ of the perturbations of the test K-G field field $\phi$ in the massless case, according to our approach the Hamiltonians take substantially different forms (see (\ref{dyn-tutte})). It will make a difference between the dynamics on the quantum background of the quantum perturbations  of the clock field $T$ on the one hand, and the dynamics of the quantum test scalar field $\phi$ on the other hand.}

{}{Our results provide a good starting point to understanding and calculating effects that quantum cosmological spacetime in the background has on the quantum perturbations of the metric tensor and the K-G field, specifically in the case of the K-G field which does not vanish in the background
(in the zeroth order).}

\acknowledgements
We thank Wojciech Kaminski for help in defining the physical Hamiltonian in an exact way consistent with the unperturbed theory. This work was partially supported by the grant of Polish Ministerstwo Nauki i Szkolnictwa Wy\.{z}szego nr N N202 104838 and by the grant of Polish Narodowe Centrum Nauki nr 2011/02/A/ST2/00300. JP's contribution is a part of the project \emph{International PhD Studies in Fundamental Problems of Quantum Gravity and Quantum Field Theory} of Foundation for Polish Science, cofinanced by the program IE OP 2007-2013 within European Regional Development Fund.
\appendix
\section{Useful formulae for the derivation of the physical Hamiltonian}
In this appendix, we collect some nontrivial formula used in the main text for the computation of the physical Hamiltonian.
		\subsection{Expansion of the curvature}
The biggest trouble with (\ref{dyn-hargument}) is due to the curvature terms. In this section we expand $R^{(3)}$, exposing $\delta^{(1)} R^{(3)}$ and $\delta^{(2)} R^{(3)}$ in terms of the linear perturbations. The next section is dedicated to the expansion of other (less) troublesome quantities, namely the determinant of the spatial metric.

The starting point is formula (\ref{prop:curvature1})
\begin{align} \label{ci-curva}
R^{(3)} = q^{ab} \left(\partial_c \Gamma^{\ c}_{a \ b} - \partial_b \Gamma^{\ c}_{a \ c} + \Gamma^{\ d}_{a \ b} \Gamma^{\ c}_{c \ d} - \Gamma^{\ d}_{a \ c} \Gamma^{\ c}_{b \ d}\right)
\end{align}
We know that Christoffel symbols always involve derivatives of the metric, so they are at least of 1st order. Thus, the last two terms are of 2nd order themselves, so they are contracted by $q^{ab}_{(0)}$, and contribute to $\delta^{(2)} R^{(3)}$. On the other hand, the first two terms contain parts of the 1st order, so they should in general be contracted with $(q^{ab}_{(0)} + \delta^{(1)} q^{ab})$. To find what the perturbation $\delta^{(1)} q^{ab}$ is, we use the definition of inverse metric, $q^{ab} q_{bc} = \delta^a_c$. From this, it follows that
\begin{align} \label{invq-expC}
\delta^a_c & = (q^{ab}_{(0)} + \delta^{(1)} q^{ab}) (q^{(0)}_{bc} + \delta q_{bc}) = q^{ab}_{(0)} q^{(0)}_{bc} + q^{ab}_{(0)} \delta q_{bc} + \delta^{(1)} q^{ab} q^{(0)}_{bc} = \notag
\\
& = \delta^a_c + q^{ab}_{(0)} \delta q_{bc} + \delta^{(1)} q^{ab} q^{(0)}_{bc}
\end{align}
up to 1st order. So, by contracting both sides with $q_{(0)}^{cd}$ we see that
\begin{align} \label{invq-exp}
\delta^{(1)} q^{ab} = -q^{ac}_{(0)} q_{(0)}^{bd} \delta q_{cd}
\end{align}
Thus, one should be not fooled in thinking that $\delta^{(1)} q^{ab}$ is simply $\delta q_{ab}$ with its indices raised via the background metric; indeed, it is almost like that, except that there is a minus sign in front of it! Knowing (\ref{invq-exp}), we can rewrite (\ref{ci-curva}) as
\begin{align} \label{ci-curvaSPLIT}
R^{(3)} = (q^{ab}_{(0)} - q^{ae}_{(0)} q_{(0)}^{bf} \delta q_{ef}) \partial_c \Gamma^{\ c}_{a \ b} - (q^{ab}_{(0)} - q^{ae}_{(0)} q_{(0)}^{bf} \delta q_{ef}) \partial_b \Gamma^{\ c}_{a \ c} + q^{ab}_{(0)} \Gamma^{\ d}_{a \ b} \Gamma^{\ c}_{c \ d} - q^{ab}_{(0)} \Gamma^{\ d}_{a \ c} \Gamma^{\ c}_{b \ d}
\end{align}
Now, inevitably, we need to take a look at the Christoffel symbols themselves. We have
\begin{align} \label{ci-christ}
\Gamma^{\ c}_{a \ b} & = \dfrac{1}{2} q^{cd} \left(\partial_a q_{bd} + \partial_b q_{ad} - \partial_d q_{ab}\right) = \dfrac{1}{2} (q^{cd}_{(0)} + \delta^{(1)} q^{cd}) \left(\partial_a \delta q_{bd} + \partial_b \delta q_{ad} - \partial_d \delta q_{ab}\right) = \notag
\\
& = \dfrac{1}{2} (q^{cd}_{(0)} - q^{ce}_{(0)} q_{(0)}^{df} \delta q_{ef}) \left(\partial_a \delta q_{bd} + \partial_b \delta q_{ad} - \partial_d \delta q_{ab}\right) = \notag
\\
& = \dfrac{e^{-2\alpha}}{2} \delta^{cd} \left(\partial_a \delta q_{bd} + \partial_b \delta q_{ad} - \partial_d \delta q_{ab}\right) - \dfrac{e^{-4\alpha}}{2} \delta^{ce} \delta^{df} \delta q_{ef} \left(\partial_a \delta q_{bd} + \partial_b \delta q_{ad} - \partial_d \delta q_{ab}\right)
\end{align}
having used the explicit form of $q_{(0)}^{ab}$. From this, we can compute the different objects we need:
\begin{align} \label{ci-objects}
\begin{array}{ll}
\partial_c \Gamma^{\ c}_{a \ b} & = \dfrac{e^{-2\alpha}}{2} \left(\partial_a \partial^d \delta q_{bd} + \partial_b \partial^d \delta q_{ad} - \partial^d \partial_d \delta q_{ab}\right) -
\\
& - \dfrac{e^{-4\alpha}}{2} (\partial^e \delta q_{ef}) \left(\delta^{df} \partial_a \delta q_{bd} + \delta^{df} \partial_b \delta q_{ad} - \partial^f \delta q_{ab}\right) - \dfrac{e^{-4\alpha}}{2} \delta q_{ef} \left(\delta^{df} \partial_a \partial^e \delta q_{bd} + \delta^{df} \partial_b \partial^e \delta q_{ad} - \partial^e \partial^f \delta q_{ab}\right)
\\
\\
\partial_b \Gamma^{\ c}_{a \ c} & = \dfrac{e^{-2\alpha}}{2} \delta^{cd} \partial_a \partial_b \delta q_{cd} - \dfrac{e^{-4\alpha}}{2} \delta^{ce} \delta^{df} (\partial_a \delta q_{cd}) (\partial_b \delta q_{ef}) - \dfrac{e^{-4\alpha}}{2} \delta q_{ef} \delta^{ce} \delta^{df} \partial_a \partial_b \delta q_{cd}
\\
\\
\Gamma^{\ d}_{a \ b} \Gamma^{\ c}_{c \ d} & = \dfrac{e^{-4\alpha}}{4} \delta^{cd} \partial_a \delta q_{be} \partial^e \delta q_{cd} + \dfrac{e^{-4\alpha}}{4} \delta^{cd} \partial_b \delta q_{ae} \partial^e \delta q_{cd} - \dfrac{e^{-4\alpha}}{4} \delta^{cd} \partial_e \delta q_{ab} \partial^e \delta q_{cd}
\\
\\
\Gamma^{\ d}_{a \ c} \Gamma^{\ c}_{b \ d} & = \dfrac{e^{-4\alpha}}{4} \delta^{ce} \delta^{df} \partial_a \delta q_{cd} \partial_b \delta q_{ef} + \dfrac{e^{-4\alpha}}{2} \partial^d \delta q_{ac} \partial^c \delta q_{bd} - \dfrac{e^{-4\alpha}}{2} \delta^{cd} \partial_e \delta q_{ac} \partial^e \delta q_{bd}
\end{array}
\end{align}
Now, we plug these in (\ref{ci-curvaSPLIT}), and retain only the terms up to 2nd order. The result is
\begin{align} \label{ci-curvaDONE}
R^{(3)} = \delta^{(1)} R^{(3)} + \delta^{(2)} R^{(2)}
\end{align}
where
\begin{align} \label{ci-curvaPARTS}
\delta^{(1)} R^{(3)} & = e^{-4\alpha} \partial^a \partial^d \delta q_{ad} - e^{-4\alpha} \delta^{ab} \partial_e \partial^e \delta q_{ab}
\\ \notag
\\
\delta^{(2)} R^{(3)} & = - 2e^{-6\alpha} \delta^{ab} \delta q_{bc} \partial^c \partial^d \delta q_{da} + e^{-6\alpha} \delta^{ac} \delta^{bd} \delta q_{ab} \partial_e \partial^e \delta q_{cd} + e^{-6\alpha} \delta q_{ab} (\delta^{cd} \partial^a \partial^b \delta q_{cd}) +
\\
& + e^{-6\alpha} (\partial^a \delta q_{ab}) (\delta^{cd} \partial^b \delta q_{cd}) - \dfrac{e^{-6\alpha}}{4} (\delta^{ab} \partial_e \delta q_{ab}) (\delta^{cd} \partial^e \delta q_{cd}) - \notag
\\
& - \dfrac{e^{-6\alpha}}{2} \delta^{ab} (\partial^d \delta q_{ac}) (\partial^c \delta q_{bd}) - e^{-6\alpha}  \delta^{cd} (\partial^a \delta q_{ac}) (\partial^b \delta q_{bd}) + \dfrac{3 e^{-6\alpha}}{4} \delta^{ac} \delta^{bd} (\partial_e \delta q_{ab}) (\partial^e \delta q_{cd}) \notag
\end{align}
		\subsection{Expansion of the determinant}
Here we present the expansion of the determinant $q$, also present in (\ref{dyn-hargument}). One simply applies the definition (up to 2nd order):
\begin{align} \label{ci-detC}
3! q & = \epsilon^{abc} \epsilon^{def} q_{ad} q_{be} q_{cf} = \epsilon^{abc} \epsilon^{def} (q^{(0)}_{ad} + \delta q_{ad}) (q^{(0)}_{be} + \delta q_{be}) (q^{(0)}_{cf} + \delta q_{cf}) = \notag
\\
& = 3! q^{(0)} + \epsilon^{abc} \epsilon^{def} (\delta q_{ad} q^{(0)}_{be} q^{(0)}_{cf} + q^{(0)}_{ad} \delta q_{be} q^{(0)}_{cf} + q^{(0)}_{ad} q^{(0)}_{be} \delta q_{cf}) + \notag
\\
& + \epsilon^{abc} \epsilon^{def} (q^{(0)}_{ad} \delta q_{be} \delta q_{cf} + \delta q_{ad} q^{(0)}_{be} \delta q_{cf} + \delta q_{ad} \delta q_{be} q^{(0)}_{cf}) = \notag
\\
& = 3! e^{6\alpha} + e^{4\alpha} \epsilon^{abc} \epsilon^{def} (\delta q_{ad} \delta_{be} \delta_{cf} + \delta_{ad} \delta q_{be} \delta_{cf} + \delta_{ad} \delta_{be} \delta q_{cf}) + \notag
\\
& + e^{2\alpha} \epsilon^{abc} \epsilon^{def} (\delta_{ad} \delta q_{be} \delta q_{cf} + \delta q_{ad} \delta_{be} \delta q_{cf} + \delta q_{ad} \delta q_{be} \delta_{cf}) = \notag
\\
& = 3! e^{6\alpha} + 2 e^{4\alpha} (\delta^{ad} \delta q_{ad} + \delta^{be} \delta q_{be} + \delta^{cf} \delta q_{cf}) + \notag
\\
& + e^{2\alpha} ((\delta^{be} \delta^{cf} - \delta^{bf} \delta^{ce}) \delta q_{be} \delta q_{cf} + (\delta^{ad} \delta^{cf} - \delta^{af} \delta^{cd}) \delta q_{ad} \delta q_{cf} + (\delta^{ad} \delta^{be} - \delta^{ae} \delta^{bd}) \delta q_{ad} \delta q_{be}) = \notag
\\
& = 3! e^{6\alpha} + 6 e^{4\alpha} \delta^{ab} \delta q_{ab} + 3 e^{2\alpha} (\delta^{ab} \delta q_{ab} \delta^{cd} \delta q_{cd} - \delta^{ab} \delta^{cd} \delta q_{ac} \delta q_{bd})
\end{align}
having used the well-known relations between $\epsilon$ and $\delta$. So we can write
\begin{align} \label{ci-det}
q = q^{(0)} + \delta^{(1)} q + \delta^{(2)} q = e^{6\alpha} + e^{4\alpha} \delta^{ab} \delta q_{ab} + \dfrac{e^{2\alpha}}{2} (\delta^{ab} \delta q_{ab} \delta^{cd} \delta q_{cd} - \delta^{ab} \delta^{cd} \delta q_{ac} \delta q_{bd})
\end{align}
		\subsection{Fourier mode-expansion}
Here, we explain how to get to equation (\ref{dyn-HP2}) via real Fourier transform. Since $h^{(2)}_\text{phys}$ is of second order in the perturbations, upon plugging the expansions (\ref{pertKexp}) in it, we will obtain something which comprises terms of the form
\begin{align} \label{genE1}
G_{abcd} = & \int d^3x \left[\delta \breve{f}_{ab}(0) + \dfrac{1}{\sqrt{2}} \sum_{k \in \mathcal{L}_+} \left(\delta \breve{f}_{ab}(k) (e^{ik \cdot x} + e^{-ik \cdot x}) + i \delta \breve{f}_{ab}(-k) (e^{ik \cdot x} - e^{-ik \cdot x})\right)\right] \times \notag
\\
& \times \left[\delta \breve{g}_{cd}(0) + \dfrac{1}{\sqrt{2}} \sum_{k' \in \mathcal{L}_+} \left(\delta \breve{g}_{cd}(k') (e^{ik' \cdot x} + e^{-ik' \cdot x}) + i \delta \breve{g}_{cd}(-k') (e^{ik' \cdot x} - e^{-ik' \cdot x})\right)\right]
\end{align}
These can be seen to reduce to
\begin{align} \label{genE2}
G_{abcd} & = \delta \breve{f}_{ab}(0) \delta \breve{g}_{cd}(0) \int d^3x + \delta \breve{f}_{ab}(0) \sqrt{2} \sum_{k \in \mathcal{L}_+} \delta \breve{g}_{cd}(k) \delta_{k, 0} + \delta \breve{g}_{cd}(0) \sqrt{2} \sum_{k \in \mathcal{L}_+} \delta \breve{f}_{ab}(k) \delta_{k, 0} + \notag
\\
& + \sum_{k, k' \in \mathcal{L}_+} \left(\delta \breve{f}_{ab}(k) \delta \breve{g}_{cd}(k') (\delta_{k, -k'} + \delta_{k, k'}) - \delta \breve{f}_{ab}(-k) \delta \breve{g}_{cd}(-k') (\delta_{k, -k'} - \delta_{k, k'})\right)\end{align}
having used the fact that
\begin{align} \label{genE3}
\int d^3x e^{i (k - k') \cdot x} = \delta_{k, k'}
\end{align}
But since $k, k'$ only take values on $\mathcal{L}_+$, then both $\delta_{k, -k'}$ and $\delta_{k', 0}$ vainsh when we perform the sum over $k'$. In other words, $G_{abcd}$ is finally reduced to
\begin{align} \label{genE4}
G_{abcd} = \delta \breve{f}_{ab}(0) \delta \breve{g}_{cd}(0) + \sum_{k \in \mathcal{L}_+} \left(\delta \breve{f}_{ab}(k) \delta \breve{g}_{cd}(k) + \delta \breve{f}_{ab}(-k) \delta \breve{g}_{cd}(-k)\right)
\end{align}
Thus, we are reduced to a sum over all $k$. In the case there are derivatives the result is the same: so, to expand in modes $k$, we first separate the $k = 0$ mode and then simply write a sum of decoupled terms over $k \in \mathcal{L} - \{0\}$, each of which perfectly resembles the corresponding one on coordinate space (with the difference that $\partial_a$ is replaced with $k_a$).\footnote{
The only non-trivial point is to remember to put a minus sign in the case both derivatives act on the same perturbation. One expects this because of the $i^2$ factor. If this is not enough to convince the reader, we suggest he repeats the computation done above replacing $G_{abcd} = \delta f_{ab} \delta g_{cd}$ with $G_{abcdef} = \delta f_{ab} \partial_e \partial_f \delta g_{cd}$.
}
Applying these rules, we indeed recover equation (\ref{dyn-HP2}) in a reasonable amount of time.
\end{document}